\documentclass[singlecolumn,10pt]{asme2ej}

\usepackage{epsfig} %% for loading postscript figures

\usepackage{amsmath}

\usepackage[final]{changes} 

\usepackage{bm}% bold math
\usepackage{algorithm}  
\usepackage{algpseudocode}  

\usepackage{lastpage}
\usepackage{fancyhdr}
\usepackage{setspace}

\title{Constructing rigid-foldable generalized Miura-ori tessellations for curved surfaces}

\author{Yucai Hu\\	    
        Department of Mechanics and Aerospace Engineering\\
        Southern University of Science and Technology\\
        Shenzhen, 518055, China\\
  		Email: huyc@sustech.edu.cn
        \\
        \\
		{\tensfb Yexin Zhou}
        \affiliation{
        	Faculty of Science, Harbin Institute of Technology\\
        	Shenzhen, 518055, China\\
        	Email: zhouyexin@hit.edu.cn
		}
}

\author{Haiyi Liang \thanks{corresponding author.} 
    \affiliation{ 
	    CAS Key Laboratory of Mechanical Behavior and Design of Materials\\
	    Department of Modern Mechanics\\
	    University of Science and Technology of China\\
	    Hefei, Anhui 230026, China\\
	    IAT-Chungu Joint Laboratory for Additive Manufacturing\\
	    Anhui Chungu 3D printing Institute of Intelligent Equipment and Industrial Technology\\
	    Wuhu, Anhui 241200, China\\
	    Email: hyliang@ustc.edu.cn
    }
}

\begin{document}

\maketitle    

%%%%%%%%%%%%%%%%%%%%%%%%%%%%%%%%%%%%%%%%%%%%%%%%%%%%%%%%%%%%%%%%%%%%%%
%\doublespacing

\begin{abstract}
{\it{   Origami has shown the potential to approximate three-dimensional curved
		surfaces by folding through designed crease patterns on flat materials.
		The Miura-ori tessellation is a widely used pattern in engineering
		and tiles the plane when partially folded. Based on constrained optimization,
		this paper presents the construction of generalized Miura-ori patterns
		that can approximate three-dimensional parametric surfaces of varying
		curvatures while preserving the inherent properties of the standard
		Miura-ori, including developability, flat-foldability and rigid-foldability.
		An initial configuration is constructed by tiling the target surface
		with triangulated Miura-like unit cells and used as the initial guess
		for the optimization. For approximation of a single target surface,
		a portion of the vertexes on the one side is attached to the target
		surface; for fitting of two target surfaces, a portion of vertexes
		on the other side is also attached to the second target surface. The
		parametric coordinates are adopted as the unknown variables for the
		vertexes on the target surfaces whilst the Cartesian coordinates are
		the unknowns for the other vertexes. The constructed generalized Miura-ori
		tessellations can be rigidly folded from the flat state to the target
		state with a single degree of freedom.}
}
\end{abstract} 

%%%%%%%%%%%%%%%%%%%%%%%%%%%%%%%%%%%%%%%%%%%%%%%%%%%%%%%%%%%%%%%%%%%%%%
\section{Introduction}

Origami has emerged as a technology to construct three-dimensional
(3D) folded structures by folding a piece of flat material. The folded
configurations that can be attained depend on the crease pattern which
consists of mountain/valley creases and vertexes. Among the various
crease patterns, the standard Miura-ori pattern is probably the most
well-studied and widely used pattern in engineering. It was proposed
for the packaging and deployment of large membranes in space \cite{miura1970proposition,miura1985method};
the pattern and its variants have found applications as the foldcores
of sandwich structures \cite{miura1972zeta,klett2011designing,heimbs2013foldcore,ma2018origami}.
Recently, unique properties of origami structures, such as negative
Poisson's ratio and multi-stability, have been revealed by researchers
and the Miura-ori tessellation has inspired the design of mechanical
and acoustic metamaterials, see \cite{wei2013geometric,schenk2013geometry,silverberg2014using,pratapa2018bloch}
among others.

\begin{figure}
	\begin{centering}
		\includegraphics[width=8cm]{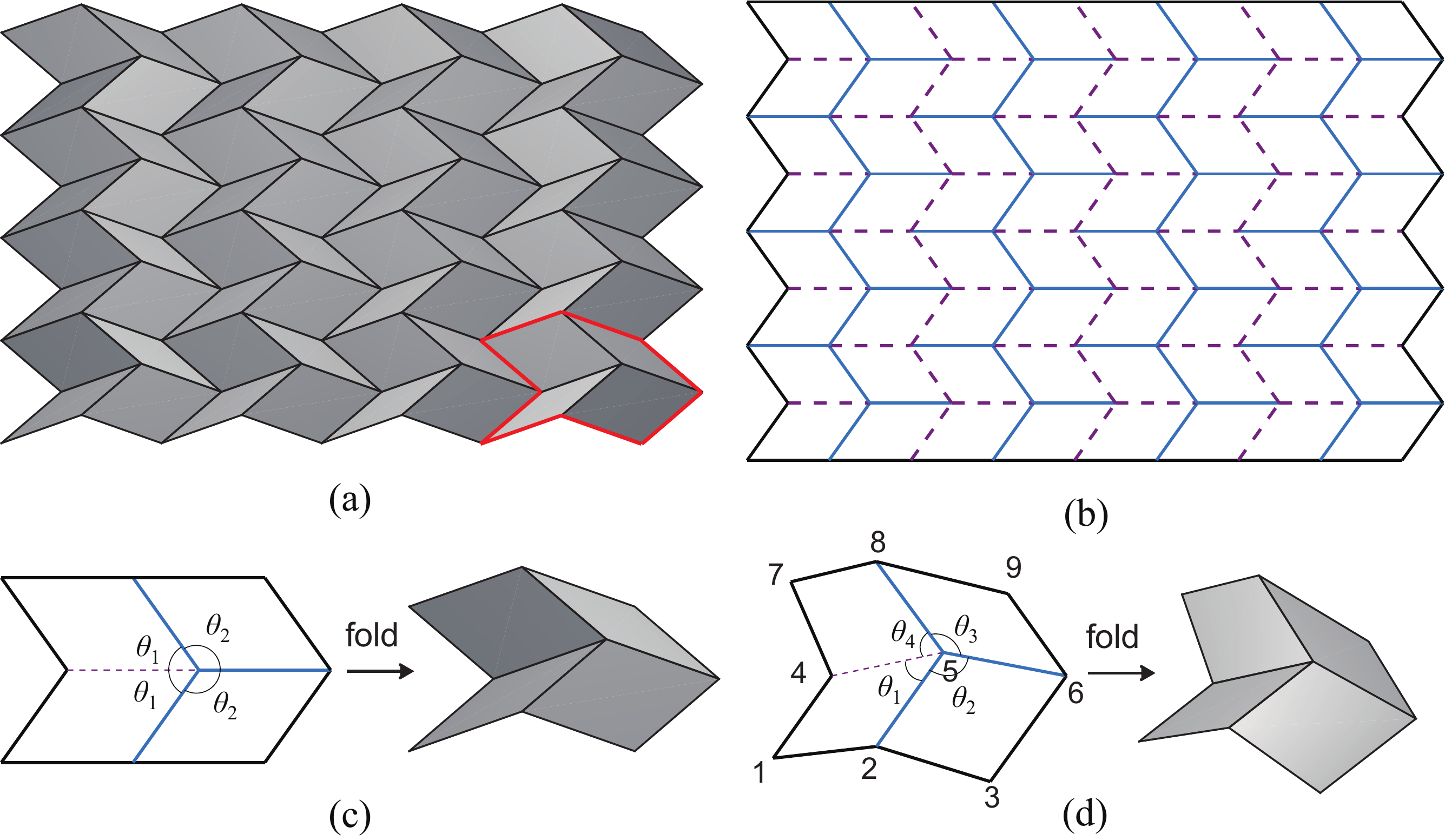}
		\par\end{centering}
	\caption{\label{fig:standared_Miura}(a) The Miura-ori tessellation with 4$\times$4
		unit cells in its partially folded state; (b) the associated crease
		pattern where the mountain and valley creases are indicated by 
		solid and dashed lines, respectively; (c) a standard Miura
		cell in which $\theta_{i}$ are the sector angles between two creases;
		(d) the generalized Miura cell with four irregular quadrilaterals;
		the sector angles are required to satisfy $\theta_{1}+\theta_{2}+\theta_{3}+\theta_{4}=2\pi$
		for developability and $\theta_{1}+\theta_{3}=\theta_{2}+\theta_{4}=\pi$
		for flat-foldability at the central vertex.}
\end{figure}

The standard Miura-ori tessellation is made up of a number of identical
unit cells, or Miura cells, and each cell consists of four congruent
parallelograms, see Fig.\ref{fig:standared_Miura}(a), (b) and (c).
Each interior vertex is of degree-4 with one pair of creases symmetric
with respect to the other collinear pair. For stiff facets connected
by soft creases, it would be much harder to bend the facets than to
fold along the creases, thus the morphing of the origami structures
can be described by the rigid origami model in which all the facets
are assumed rigid and only folding along the creases are allowed.
Treated as the rigid origami, the Miura-ori tessellation constitutes
a mechanism which folds with a single degree of freedom (DOF) before
all the facets collapse into the same plane. In all the partially
folded states, the Miura-ori tessellation can only exhibit in-plane
motion and tiles the plane. To approximate 3D curved surfaces, variations
in the edge length and sector angles of the cells should be involved,
see Fig.\ref{fig:standared_Miura}(d) for an illustration of a generalized
Miura cell. However, for a general quadrilateral mesh, rigid-foldability
is nontrivial since the compatibility between all the rigid facets
during the folding process can lead to an overconstrained mechanism
\cite{tachi2009generalization,dieleman2019jigsaw}. It has been shown
by Tachi that standard Miura-ori pattern can be perturbed into general
ones which satisfy the developability, flat-foldability and other
constraints \cite{tachi2009generalization,tachi2010freeform_quad}.
The resulting generalized patterns can approximate some curved surfaces
when partially folded. Most importantly, it has been proven that if
an intermediate folded state exists for a quadrilateral mesh consisting
of developable and flat-foldable vertexes, the pattern is rigid-foldable
limited only by the avoidance of self-penetrations \cite{tachi2009generalization,lang2017twists}.
Based on the findings, Tachi has developed the ``Freeform origami''
software which can modify the standard Miura-ori tessellation (and
other tessellations) into some curved structure by interactively moving
the vertexes and projecting the perturbation into the constraint space
\cite{tachi2010freeform,tachi2010freeform_quad}. However, it is difficult
for the software to tackle inverse design problems since constraints
regarding approximating target surfaces are not considered yet. By
altering a single characteristic of the Miura-ori cell, Gattas et
al. have investigated the parameterizations of first-level Miura-ori
derivative patterns \cite{gattas2013miura}. Lang and Howell proposed
a method to construct rigid-foldable quadrilateral meshes based on
the inherent relations between the sector and fold angles of flat-foldable
vertexes \cite{lang2018rigidly}. Recently, Dieleman et al. developed
a systematic approach by first identifying a number of rigid-foldable
motifs as jigsaw puzzle pieces and then fitting them together to obtain
large rigid-foldable crease patterns \cite{dieleman2019jigsaw}.

Several approaches have been developed for the design of cylindrical
and axisymmetric origami structures \cite{zhou2015design,wang2016folding,song2017design,hu2019design}.
Due to the symmetry of the target cylindrical/axisymmetric surfaces,
the crease patterns observe collinear creases which are parallelly
(radially) spaced for the cylindrical (axisymmetric) cases. Though
restrictive in the designed geometry, the approaches in the references
\cite{zhou2015design,wang2016folding,song2017design,hu2019design}
can be used to design folded structures which fit two given target
surfaces. Dudte et al. have proposed an optimization-based algorithm
to find the intermediate folded form which approximates general curved
surfaces \cite{dudte2016programming}. In this algorithm, the target
surface is first tiled with a Miura-like tessellation. To ensure the
approximation, the four corner vertexes of each cell (vertexes 1,
3 ,7 and 9 of each cell shown in Fig.\ref{fig:standared_Miura}(d))
are fixed as the initial guess whilst the positions of the other vertexes
are searched by the optimization method to meet the constraints, including
the planarity of all the quadrilaterals and the developability at
all the interior vertexes. However, it was reported that the algorithm
fails to find the folded form that satisfies the flat-foldability
condition for general curved surfaces. Due to the aforementioned theorem
by Tachi \cite{tachi2009generalization}, the patterns constructed
by Dudte et al. are not rigid-foldable, i.e., the quadrilateral facets
will be bent along the diagonals when it is folded from the flat state
to the target state. To approach the flat-foldability (and thus the
rigid-foldability), the flat-foldability constraint is replaced by
inequalities with auxiliary tolerance \cite{dudte2016programming}.
However, this remedy can only reduce the bending energy of the quadrilateral
whilst the crease patterns consist of triangular facets during the
folding process. While introducing diagonal creases enlarges the space
of rigid-foldable configurations, extra efforts are required when
folding the flat pattern into the target form due to the introduced
DOFs and the complex interaction between the energy potential and
the rigid origami \cite{waitukaitis2015origami}. The failure in satisfying
the flat-foldability condition may be caused by the fact that all
the four corner vertexes of each cell are fixed during the optimization
which reduces the number of variables in the algorithm. In fact, the
position of corner vertexes should also be rearranged to enlarge the
searching space such that flat- and rigid-foldable crease pattern
may be found for a curved target surface.

In this paper, based on the constrained optimization, we present the
construction of generalized Miura-ori tessellation for 3D curved parametric
surfaces which can be folded rigidly from the flat state to the target
form with a single degree of freedom. 
The construction is based on the rigid origami model, 
which ngelects the deformation of facets and creases.
The construction utilizes general
quadrilateral mesh with developable and flat-foldable interior vertexes
whilst the mountain and valley crease assignment is kept the same
as that of the Miura-ori pattern. The target surface is first tiled
with triangulated Miura-like cells and then the optimization method
is used to enforce the constraints, including the planarity of all
the quadrilaterals, developability and flat-foldability at all the
interior vertexes. To ensure the fitting, a portion of vertexes on
the one side is attached to the target surface; another portion of
vertexes on the other side is also attached to the second target surface
if a second target surface is to be approximated. For the attached
vertexes, the parametric coordinates are adopted as the unknown variables
whilst the Cartesian coordinates are the unknowns for the other vertexes. 
Several examples of fitting a single target surface and two target
surfaces are considered to demonstrate the feasibility of the algorithm.

\section{Approximate a single target surface}

\subsection{Initial tessellation construction and the constraints}

Fig.\ref{fig:standared_Miura}(c) shows the standard Miura-ori cell
made up of four congruent parallelograms. The interior vertex is of
valence four with one valley crease and three mountain creases or
vice versa. Besides, one pair of creases are symmetric with respect
to the other collinear pair. The Miura tessellation is the repeating
of the unit cells and tiles the plane in its partially folded states.
To approximate curved surfaces, the edge length and sector angles
of the cells should vary across the pattern while the basic rules
of Miura pattern, including developability, flat-foldability and rigid-foldability,
have to be fulfilled. A generalized Miura cell is illustrated in Fig.\ref{fig:standared_Miura}(d)
which consists of four quadrilaterals and the cell is developable
and flat-foldable, i.e., the sector angles around the central vertex
satisfy 
\begin{equation}
\theta_{1}+\theta_{2}+\theta_{3}+\theta_{4}=2\pi\quad\text{and}\quad\theta_{1}+\theta_{3}=\theta_{2}+\theta_{4}=\pi
\end{equation}
for developability and flat-foldability, respectively \cite{tachi2009generalization,lang2017twists}.

As pointed out by Dudte et.al \cite{dudte2016programming}, the basic
steps for constructing tessellations for curved surfaces based on
optimization are: first, generate an initial Miura-like tessellation
as an initial guess for the optimization algorithm which does not
have to satisfy the constraints; then, enforce the constraints, including
developability, flat-foldability and etc, by the constrained optimization
with appropriate objective function. Similar ideas have also been
adopted by Bhooshan to generate curved folding based on the dynamic
relaxation framework \cite{bhooshan2016interactive}.

\begin{figure*}
	\begin{centering}
		\includegraphics[width=13cm]{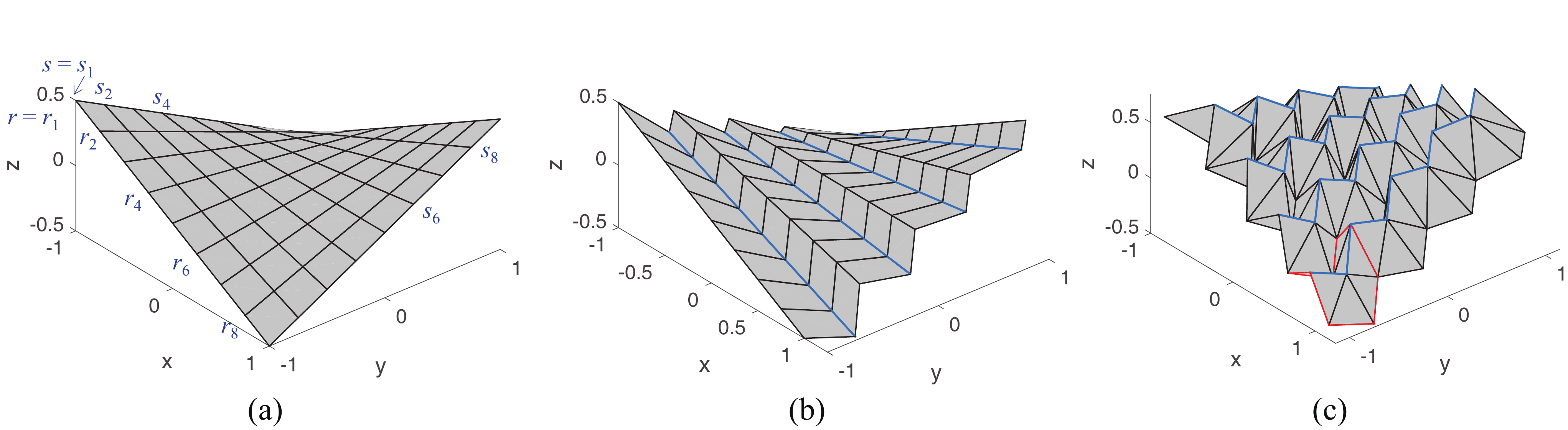}
		\par\end{centering}
	\caption{\label{fig:initial_config}
		Generation of the initial 4$\times$4 triangulated
		Miura-like tessellation for the target surface $z=xy/2$, i.e., $\textbf{X}(r,s)=\{r,s,rs/2\}$:
		(a) the base mesh formed by the parametric lines $r=r_{i}$ and $s=s_{i}$
		for $i=1,2,\cdots,9$; the $r_{i}$-coordinate lines are equispaced
		in the parametric coordinate $(\Delta r=2/8=1/4)$ and so are the
		$s_{i}$-coordinate lines; (b) the vertexes on the $s_{i}$-coordinate
		lines ($j=2,4,6,8$) are moved along the $r$-coordinate by $\Delta r$;
		(c) the vertexes on the $r_{i}$-coordinate line ($i=2,4,6,8$) are
		moved along the normal of the target surface by $1.8\Delta r$; triangulate
		the quadrilateral mesh by connecting vertex-($r_{i},s_{j}$) with
		vertex-($r_{i-1},s_{j+1}$) for $j=1,3,5,7$ and vertex-($r_{i},s_{j}$)
		with vertex-($r_{i+1},s_{j+1}$) for $j=2,4,6,8$.}
\end{figure*}

For expository purposes, we consider the target surface $z=xy/2$
with $x,y\in[-1,1]$ as an example. The target surface can be expressed
in the parametric form as $\textbf{X}(r,s)=\{r,s,rs/2\}$. Fig.\ref{fig:initial_config}
shows the construction of the initial triangulated Miura-like tessellation:
%\begin{itemize}

$\bullet$ A base mesh is first generated by the $r=r_{i}$ and $s=s_{i}$ coordinate
	lines with $r_{i}=-1+(i-1)\Delta r$ and $s_{i}=-1+(i-1)\Delta s$
	for $i=1,2,\cdots,9$ and $\Delta r=\Delta s=1/4$. In other words,
	the $r_{i}$- and $s_{i}$-coordinate lines are equispaced in the
	$r$- and $s$-coordinates, respectively.
	
$\bullet$ Then, the vertexes on the $s_{i}$-coordinate lines with $i=2,4,6$
	and 8 are moved along the $r$-coordinate by $l_{p}=\Delta r$, see
	Fig.\ref{fig:initial_config}(b).
	
$\bullet$ Finally, the vertexes on the $r_{i}$-coordinate line with $i=2,4,6$
	and 8 are moved along the normal of the target surface by $l_{h}=1.8\Delta r$;
	the quadrilateral mesh is triangulated by connecting vertex-($r_{i},s_{j}$)
	with vertex-($r_{i-1},s_{j+1}$) for $j=1,3,5,7$ and vertex-($r_{i},s_{j}$)
	with vertex-($r_{i+1},s_{j+1}$) for $j=2,4,6,8$.
	
\begin{figure}
	\begin{centering}
		\includegraphics[width=8cm]{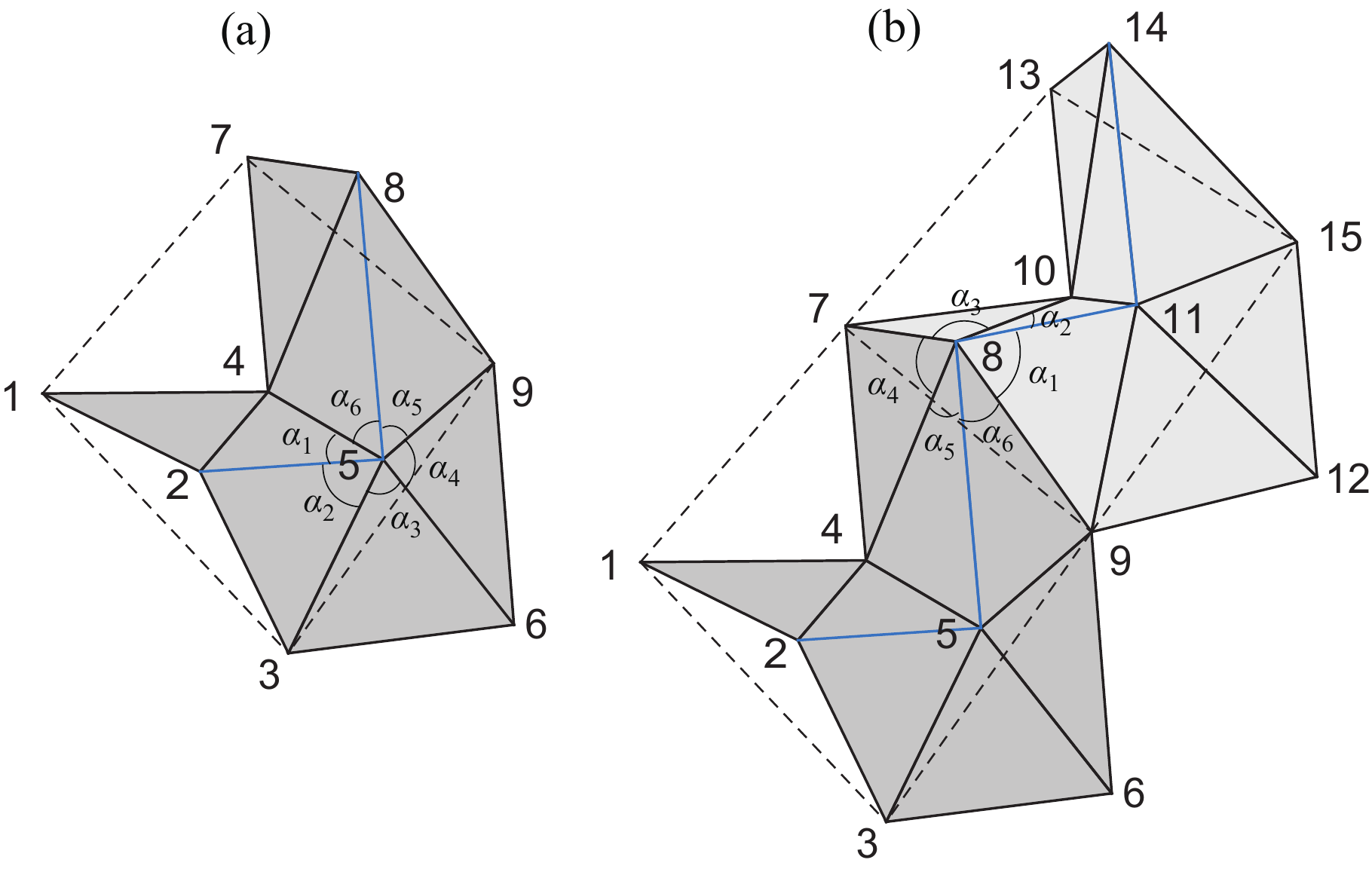}
		\par\end{centering}
	\caption{\label{fig:GM_cell} (a) A representative cell: $\alpha_{1}$ through
		$\alpha_{6}$ are the sector angles and the dashed lines indicate
		the auxiliary cell contour edges; (b) two adjacent cells. Due to the
		orientation of the quadrilateral facets and triangulation, the numbering
		of the sector angles for interior vertexes on the $s_{i}$-dinate
		lines with $i$ being even and odd follow that around vertex-5 in
		(a) and vertex-8 in (b), respectively.}
\end{figure}

%\end{itemize}
The vertexes of each quadrilateral are noncoplanar in general and
a representative triangulated cell is shown in Fig.\ref{fig:GM_cell}(a).
It should be remarked that the initial triangulated Miura-like tessellation
is not unique. By changing $l_{p}$ and $l_{h}$ (comparable to $\Delta r$
in general), different initial tessellations can be constructed. Compared
with the construction process in the reference \cite{dudte2016programming},
the current construction suits for a large class of parametric surfaces
and avoids the process of merging vertexes shared by adjacent cells.

The triangulated Miura cell in Fig.\ref{fig:GM_cell}(a) can be specified
by the coordinates of the 9 vertexes. The planarity of the quadrilateral
facets are to be restored in the optimization process. For the quadrilateral
1-2-5-4, the planarity of the quadrilateral facet requires that the
volume of the hexagon formed by the vectors $\overrightarrow{P_{1}P_{2}}$,
$\overrightarrow{P_{1}P_{4}}$ and $\overrightarrow{P_{1}P_{5}}$
equals 0, i.e.,
\begin{equation}
(\textbf{X}_{2}-\textbf{X}_{1})\times(\textbf{X}_{4}-\textbf{X}_{1})\cdot(\textbf{X}_{5}-\textbf{X}_{1})=0\label{eq:quad_planar}
\end{equation}
where $\overrightarrow{P_{i}P_{j}}$ is the vector from vertex-$i$
to vertex-$j$ and $\textbf{X}_{i}$ is the Cartesian coordinate for
the vertex-$i$. The planarity for the other quadrilaterals can be
expressed analogously. Due to the triangulation, all the inner vertexes
are of degree-6, see Fig.\ref{fig:initial_config}(c). In Fig.\ref{fig:GM_cell}(a),
the sector angles are denoted anticlockwise as $\alpha_{1}$ through
$\alpha_{6}$ around the central vertex-5. The developability condition
at vertex-5 is
\begin{equation}
\sum_{i=1}^{6}\alpha_{i}=2\pi\ .\label{eq:tri_devep}
\end{equation}
When the coplanarity of the quadrilaterals 2-3-6-5 and 5-6-9-8 are
restored, the diagonal lines 3-5 and 5-9 becomes redundant and the
vertex-5 reduces to degree-4 vertex. In the optimization process,
the flat-foldability condition at vertex-5 is
\begin{equation}
\alpha_{1}+\alpha_{4}+\alpha_{5}=\pi\ .\label{eq:tri_flat}
\end{equation}
It is clear that the identity, $\alpha_{2}+\alpha_{3}+\alpha_{6}=\pi$,
holds if Eq.(\ref{eq:tri_devep}) and Eq.(\ref{eq:tri_flat}) are
satisfied. Considering the orientation of the quadrilateral facets
and triangulation, the numbering of the sector angles in Fig.\ref{fig:GM_cell}(a)
can be used for vertexes on the $s_{i}$-coordinate lines with $i=2,4,6,8$ 
while the numbering shown in Fig.\ref{fig:GM_cell}(b) should
be adopted for the vertexes on the the $s_{\ensuremath{i}}$-coordinate
lines with $i=1,3,5,7$ such that Eq.(\ref{eq:tri_flat}) is consistent for all interior
vertexes. To facilitate the optimization algorithm, the derivatives
of the constraints in Eq.(\ref{eq:quad_planar}), (\ref{eq:tri_devep})
and (\ref{eq:tri_flat}) with respect to the variables are supplied
in the Appendix \ref{sec:append_derivative}. The triangulated Miura-like cell
will turn into a generalized Miura cell if the constraints on quadrilateral
facet planarity, developability and flat-foldability are satisfied
simultaneously.

\subsection{The variables and objective function}

To enforce a close approximation to the target surface, a portion
of the vertexes is attached to the target surface. For the cell in
Fig.\ref{fig:GM_cell}(a), it is most likely that the vertexes in
the set \{1,4,7,3,6,9\} reside on the one side while vertexes in \{2,5,8\}
on the other. Here, the corner vertexes of each cell, i.e., vertexes
\{1,7,3,9\} in Fig.\ref{fig:GM_cell}(a), are chosen to be attached
to the target surface. The attached vertexes are allowed to slide
on the target surface. For the attached vertex-$j$, it is natural
to use their Cartesian coordinates $\textbf{X}^{j}=\{x^{j},y^{j},z^{j}\}$
as the unknown variables while enforcing the attachment by satisfying
the equation of the target surface, for example in the form of $z=f(x,y)$.
This approach is still feasible when the target surface has explicit
expression regarding $x$, $y$ and $z$ coordinates. However, for
many parametric surfaces, it would be difficult to derive such a form
and numerical root finding and derivative calculations are required.
Instead, we adopt the parametric coordinates \{$r^{j},s^{j}$\} as
the unknown variables for the attached vertex-$j$, i.e., $\textbf{X}^{j}=\{x(r^{j},s^{j}),y(r^{j},s^{j}),z(r^{j},s^{j})\}$,
thus the vertex-$j$ naturally resides on the target surface. Aside
from the attached vertexes, the Cartesian coordinates are adopted
as the unknown variables for the other vertexes.

With the triangulated Miura-like tessellation as an initial guess,
the optimization algorithm is utilized to adjust the vertex positions
such that the constraints are fulfilled while minimizing appropriate
objective function. Let $N$ be the set of all the vertexes; $E_{A}$
is set of auxiliary cell contour edges for all the triangulated Miura
cells, see the dashed lines in Fig.\ref{fig:GM_cell}; $E$ be the
set of all the edges and the total number of edges is $n_{E}$. Assuming
that the initial triangulated Miura-like tessellation is a reasonable
guess, the following objective function is suggested

\begin{equation}
f=\sum_{i\in E\bigcup E_{A}}(\frac{L^{i}}{L_{0}^{i}}-1)^{2}+\sum_{j\in N}\frac{1}{(L_{c})^{2}}(\textbf{X}^{j}-\textbf{X}_{0}^{j})^{2}\label{eq:obj}
\end{equation}
where $L^{i}$ is the edge length of the current configuration whilst
$L_{0}^{i}$ is the counterpart of the initial guess; $L_{c}=\sum_{i\in E}L_{0}^{i}/n_{E}$
is the characteristic length; $\textbf{X}^{j}=\{x^{j},y^{j},z^{j}\}$
is the coordinate for the vertex-$j$ of the current configuration
whilst $\textbf{X}_{0}^{j}$ is the counterpart of the initial guess.
Furthermore, the parametric coordinates \{$r^{j},s^{j}$\} are the
unknowns for the attached vertexes, i.e., $\textbf{X}^{j}=\{x(r^{j},s^{j}),y(r^{j},s^{j}),z(r^{j},s^{j})\}$,
whilst the Cartesian coordinates $\{x^{j},y^{j},z^{j}\}$ are the
variables for the other vertexes. The first term on the left hand
of Eq.(\ref{eq:obj}) is to preserve the relative distance between
the vertexes such that no crease or boundary edge degenerates. The
second term is to ensure that the vertexes are as close as possible
to the initial guess, which also prevents the entire tessellation
from sliding on the target surface. Both terms are dimensionless such
that the objective function is scale independant.

For the tessellation with $m\times n$ triangulated Miura-like cells,
there are $2m\times2n$ quadrilaterals and each quadrilateral brings
in one planarity constraint, see Eq.(\ref{eq:quad_planar}). Besides,
there are $(2m-1)\times(2n-1)$ interior vertexes and, at each inner
vertex, Eq.(\ref{eq:tri_devep}) and Eq.(\ref{eq:tri_flat}) should
be satisfied for the local constraints on developability and flat-foldability.
Thus, the total number of constraints is 
\begin{equation}
N_{c}=4mn+2(2m-1)(2n-1)=12mn-4m-4n+2\ .\label{eq:total_cons}
\end{equation}
It is clear that the total number of vertexes is $(2m+1)(2n+1)$.
As the four corner vertexes of each cell are chosen for the attachment,
there are $(m+1)\times(n+1)$ vertexes with their parametric coordinates
as the unknowns and each parametric vertex owns 2 variables. The total
number of unknown variables and the excess DOFs are, respectively,
\begin{equation}
\begin{array}{c}
N_{f}^{s}=3(2m+1)(2n+1)-(m+1)(n+1)=11mn+5m+5n+2\\
\text{and }N_{f}^{s}-N_{c}=-mn+9(m+n)\text{ .}
\end{array}\label{eq:excess_dof_s}
\end{equation}
When $N_{f}^{s}<N_{c}$, i.e., there are more constraints than variables,
the optimization algorithm generally fails to find a generalized Miura-ori
tessellation if there were no sufficient redundancy among the constraints.
For the case of $m=n$, it is required that $m\leq18$ to ensure that
$N_{f}^{s}\geq N_{c}$.

The objective function in Eq.(\ref{eq:obj}) together with the constraints
in Eq.(\ref{eq:quad_planar}), (\ref{eq:tri_devep}) and (\ref{eq:tri_flat})
can be solved by standard optimization routines. Here, we adopt the
interior-point method implemented in the \emph{fmincon} routine of
MATLAB. To facilitate the solution, the derivatives of the objective
and constraint functions with respective to (w.r.t) the variables
should be provided for \emph{fmincon} and are discussed in the Appendix \ref{sec:append_derivative}.

\begin{figure}
	\begin{centering}
		\includegraphics[width=8cm]{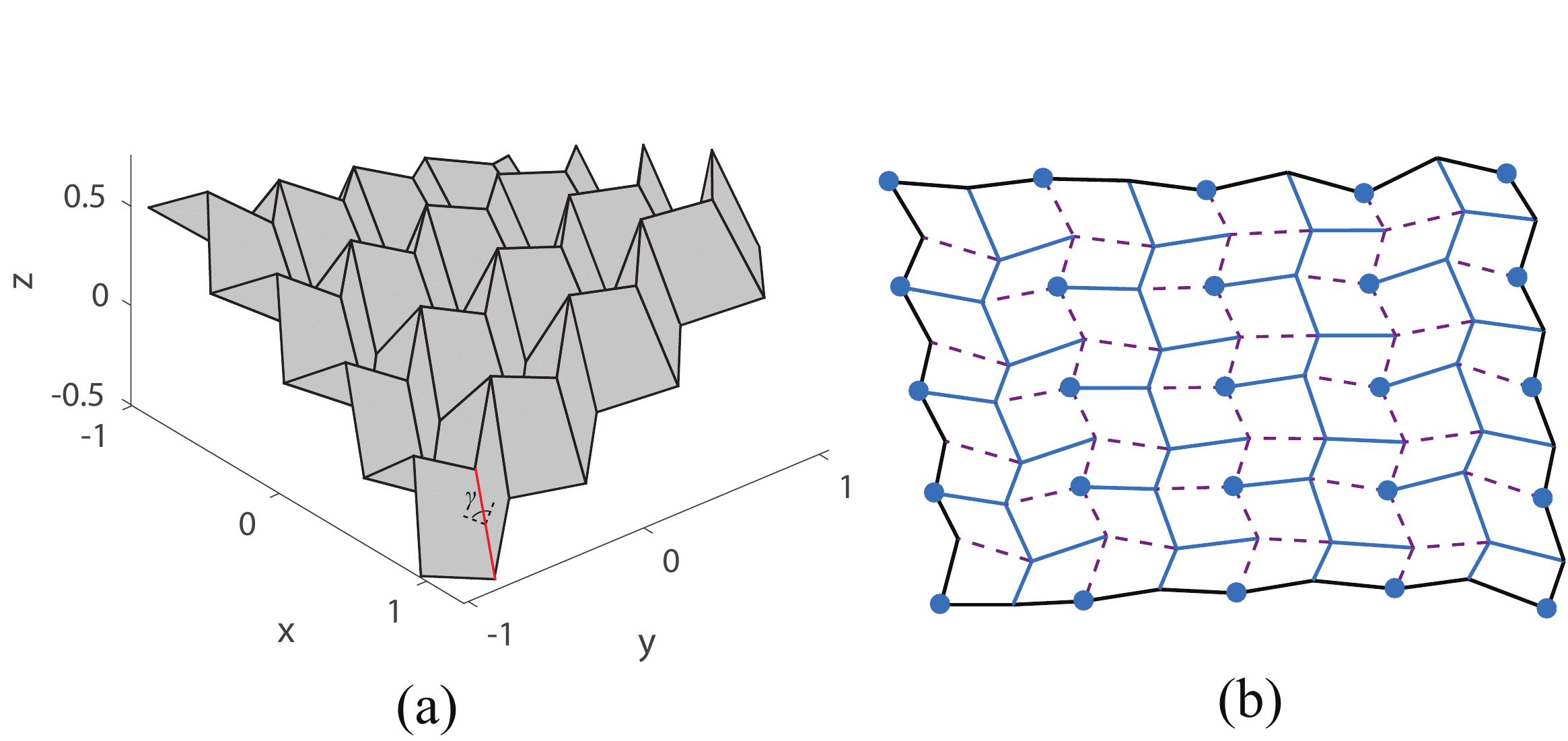}
		\par\end{centering}
	\caption{\label{fig:zxy_finial}(a) The converged result of the 4$\times$4
		Miura-ori tessellation for the target surface $z=xy/2$ and $\gamma$
		($\approx87.4^\circ$) is the dihedral angle
		at the crease; (b) the crease pattern in which the solid dots are
		the vertexes attached to the target surface when the pattern is rigidly
		folded to the target state.}
\end{figure}

\begin{figure*}
	\begin{centering}
		\includegraphics[width=13cm]{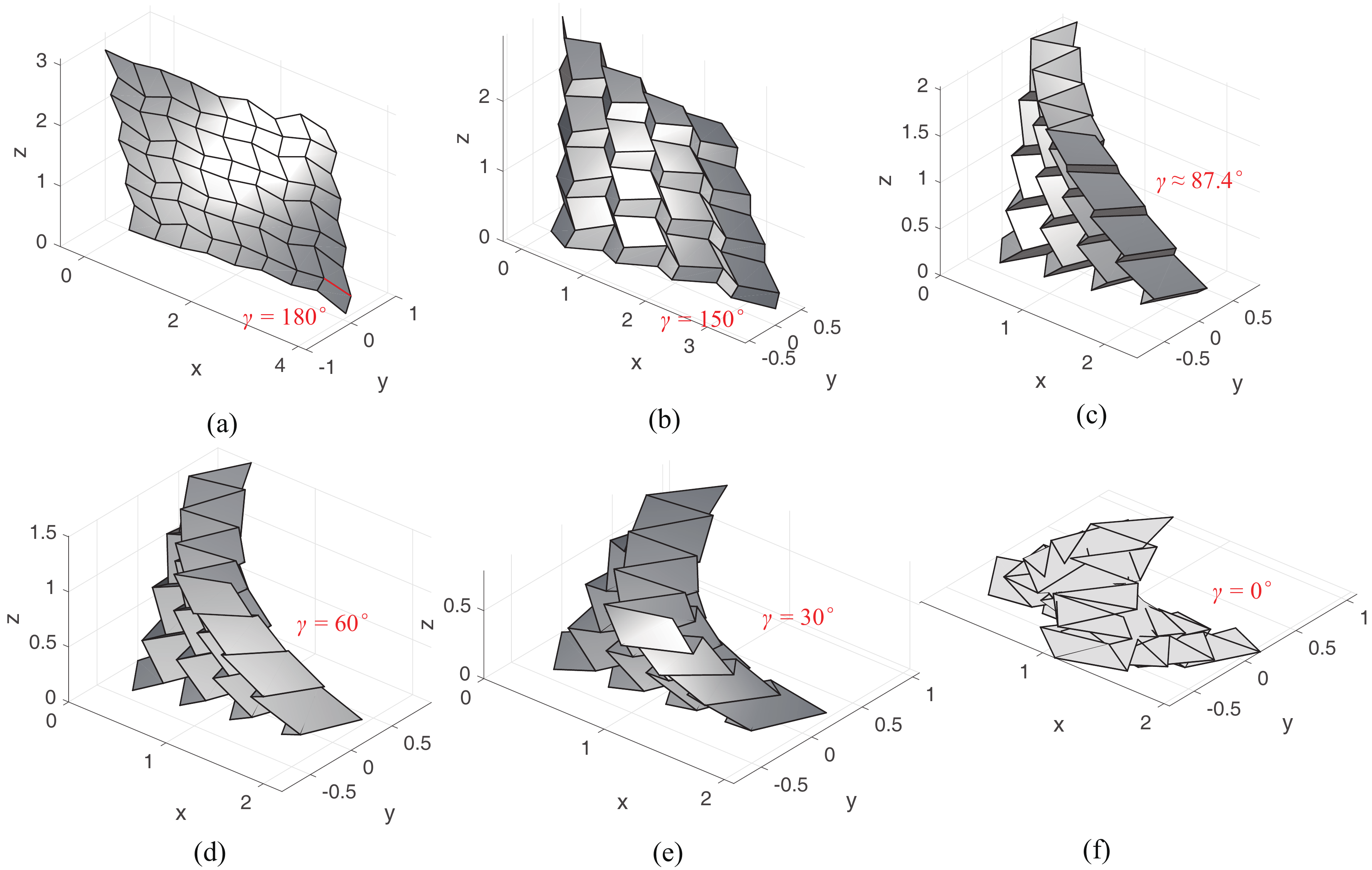}
		\par\end{centering}
	\caption{\label{fig:zxy_folding}The snapshots for the folded form specified
		by the dihedral angle $\gamma$. (a) is the flat state; (c) is the
		same configuration as that in Fig.\ref{fig:zxy_finial}(a) under a
		different view angle; (f) is the fully folded state with all facets
		being coplanar.}
\end{figure*}

For the construction of generalized Miura-ori tessellation with 4$\times$4
cells for $z=xy/2$, the \emph{fmincon} converges in 93 iterations
and the constraints are satisfied to the machine accuracy. The converged
configuration is shown in Fig.\ref{fig:zxy_finial}. It has been proved
by Tachi that if there exist a partially folded form of the quadrilateral
mesh with interior developable and flat-foldable vertexes, then the
quadrilateral mesh is rigid-foldable limited only by the penetration
avoidance (see Theorem 2 in the reference \cite{tachi2009generalization}).
Thus, the origami is rigid-foldable with a single DOF. The crease
pattern is given in Fig.\ref{fig:zxy_finial}(b) in which the vertexes
indicated by the blue dot are exactly on the target surface in the
folded form shown in Fig.\ref{fig:zxy_finial}(a). Since the origami
constitutes a single DOF mechanism, the folded state can be characterized
by any one of the dihedral angles between two adjacent facets. Here,
as indicated in Fig.\ref{fig:zxy_finial}(a), the dihedral angle $\gamma$
at the red crease is chosen which equals approximately 87.4$^\circ$
for the target state. Following the procedures in Section 4 and Appendix
A of the reference \cite{lang2018rigidly}, the dihedral angles at
all the interior creases and thus the 3D folded form can be obtained
for given $\gamma$. It is checked that no penetration occurs between
the facets, see Appendix \ref{sec:self-intersection} for detailed discussions.
Several snapshots during the folding process are shown in Fig.\ref{fig:zxy_folding}.


\subsection{Further design examples\label{subsec:signle_de}}

\begin{figure*}
	\begin{centering}
		\includegraphics[width=14cm]{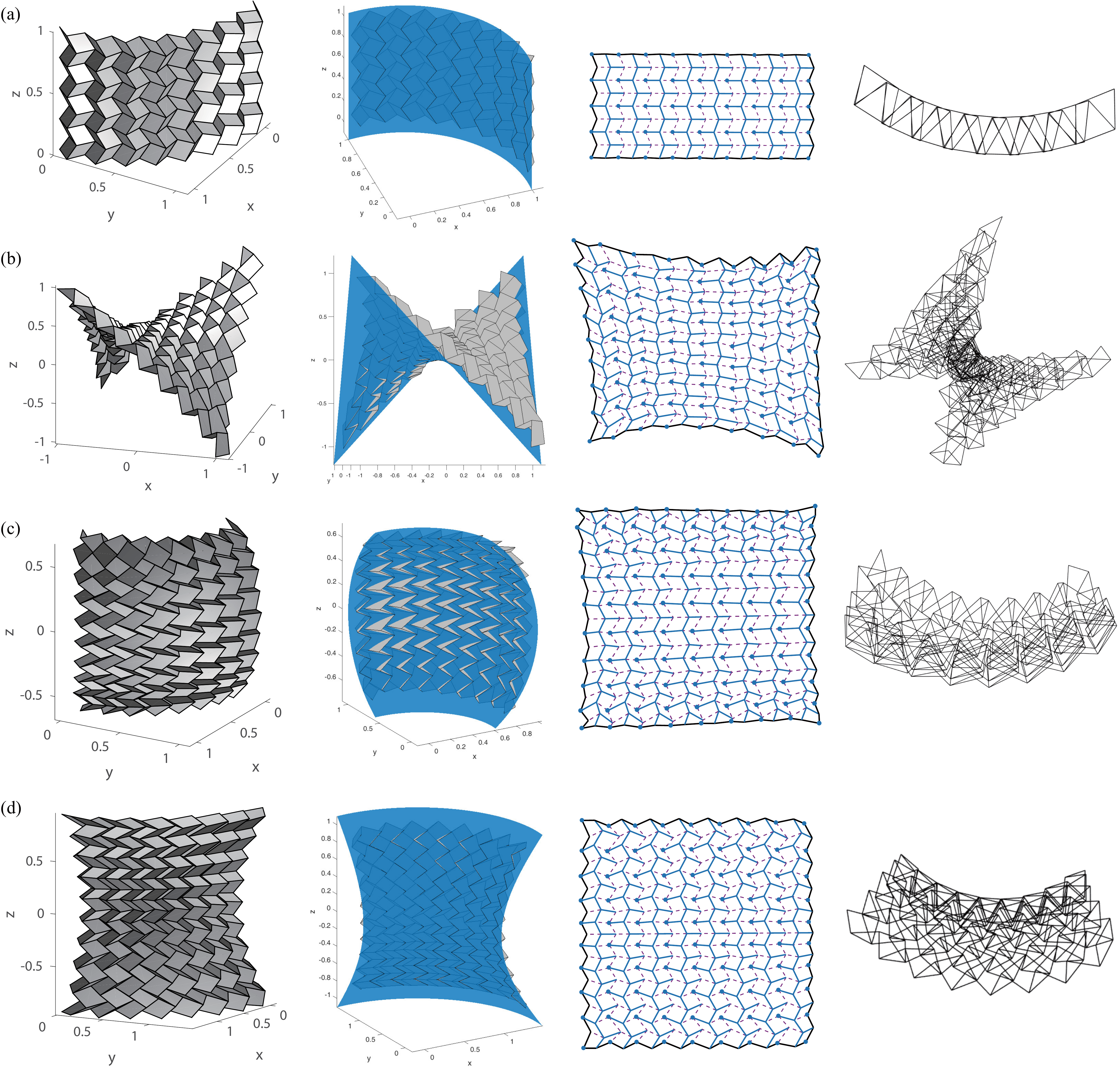}
		\par\end{centering}
	\caption{\label{fig:single_surf_eaxples}Examples of approximating a single
		target surface: the folded forms (first and second columns), the crease
		patterns (third column) and the fully folded state (fourth column),
		details are listed in Table \ref{tab:single_surf}. For figures in
		the second column, the target surface is also shown. For each
		case, the solid dots in the crease pattern in the third column are
		on the target surface when the pattern is folded to the state shown
		in the second column.}
\end{figure*}

This subsection considers several examples to further illustrate the
design algorithm which is implemented in MATLAB. The code is executed
on a desktop with Intel(R) Core(TM) i7-6700 (8 cores, 3.41Gz). The
details of the design cases are summarized in Table \ref{tab:single_surf}.
The most time consuming part of the whole construction is to solve
the optimization problem by the interior-point method of \emph{fmincon}.
As a rough reference of computational efficiency, the number of iterations
and computational time (only the solution time of the \emph{fmincon}
routine) are listed. Besides, it is checked that the constraints including
quadrilateral facet planarity, developability and flat-foldability
at the interior vertexes are fulfilled to machine accuracy (of order
less than $10^{-13}$). From the Table \ref{tab:single_surf}, it
takes more iterations to search the solution for the sphere case.
The converged folded forms, the crease pattern and the fully folded
state are, respectively, shown in the first, second and third columns
of Fig.\ref{fig:single_surf_eaxples}. It can be seen that the crease
patterns for the cases (a), (c) and (d) are roughly symmetric with
respect to the middle horizontal line.

\begin{table*}
	\caption{\label{tab:single_surf}Summary of the design examples for single target surface 
		(a) cylinder: $x^{2}+y^{2}=1$, (b) hyperbolic paraboloid:
		$z=xy$, (c) sphere: $x^{2}+y^{2}+z^{2}=1$ and (d) hyperboloid: $x^{2}+y^{2}-z^{2}=1$.}
	
	\centering{}\centerline{%
		\begin{tabular}{|c|c|c|c|c|}
			\hline 
			{\footnotesize{}cases} & {\footnotesize{}Target surface $\{x,y,z\}$} & {\footnotesize{}Cells} & {\footnotesize{}Iterations} & {\footnotesize{}Time(sec)}\tabularnewline
			\hline 
			\hline 
			{\footnotesize{}(a)} & {\footnotesize{}$\{\cos(r),\sin(r),s\}$} & {\footnotesize{}8$\times$4} & {\footnotesize{}161} & {\footnotesize{}33.3}\tabularnewline
			\hline 
			{\footnotesize{}(b)} & {\footnotesize{}$\{r,s,rs\}$} & {\footnotesize{}8$\times$8} & {\footnotesize{}148} & {\footnotesize{}145.2}\tabularnewline
			\hline 
			{\footnotesize{}(c)} & {\footnotesize{}$\{\cos(s)\cos(r),\cos(s)\sin(r),\sin(s)\}$} & {\footnotesize{}8$\times$9} & {\footnotesize{}262} & {\footnotesize{}403.6}\tabularnewline
			\hline 
			{\footnotesize{}(d)} & {\footnotesize{}$\{\sqrt{1+s^{2}}\cos(r),\sqrt{1+s^{2}}\sin(r),s\}$} & {\footnotesize{}8$\times$9} & {\footnotesize{}151} & {\footnotesize{}192.8}\tabularnewline
			\hline 
	\end{tabular}}
\end{table*}

\section{Fit two target surfaces}

\subsection{Construction process}

\begin{figure*}
	\begin{centering}
		\includegraphics[width=13cm]{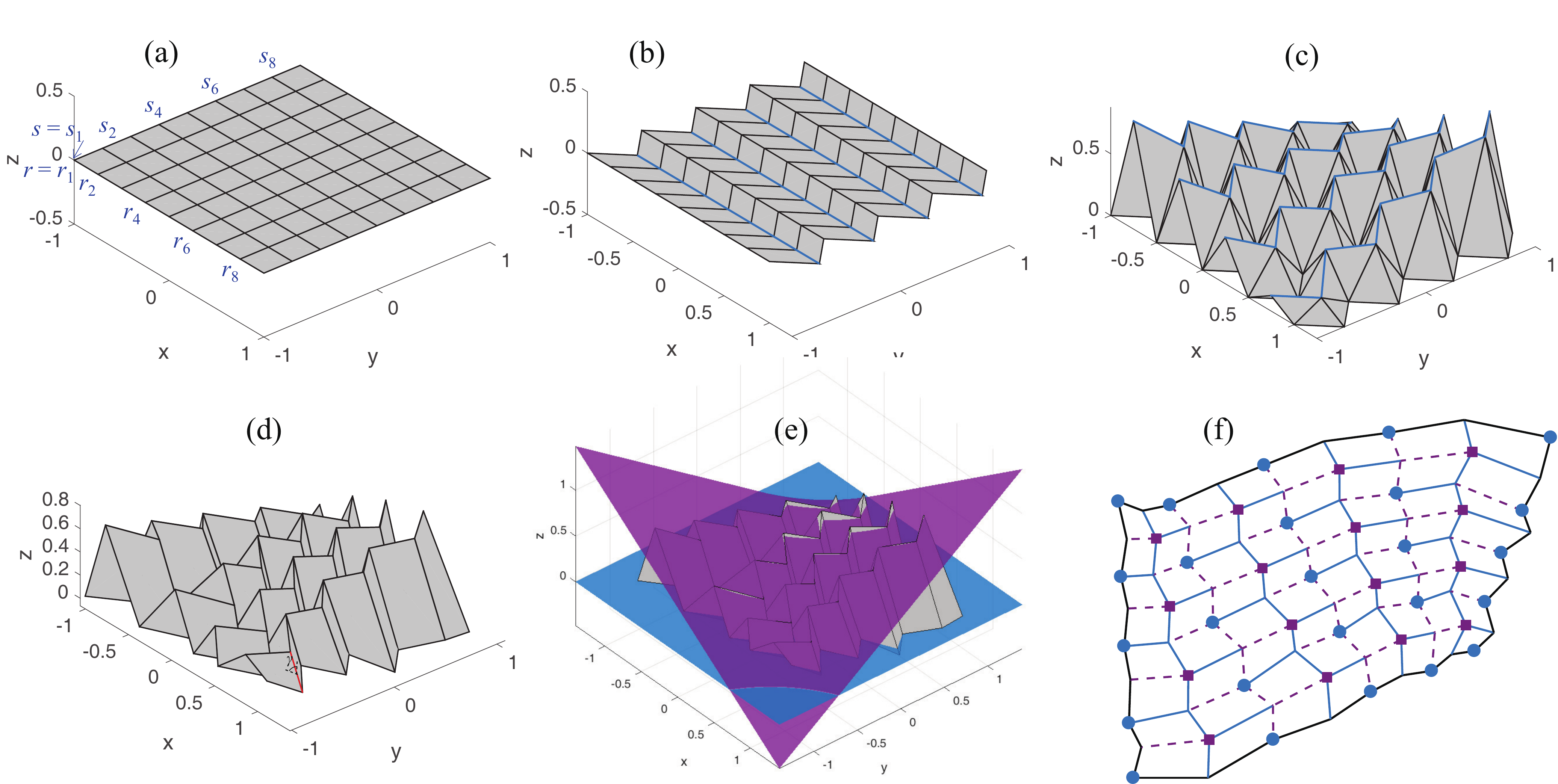}
		\par\end{centering}
	\caption{\label{fig:2surf_zxy_demo}
		For the two target surfaces $\textbf{X}_{l}=\{r,s,0\}$
		and $\textbf{X}_{u}=\{r,s,(1+rs)/2\}$, (a), (b) and (c) illustrate
		the generation of the initial 4$\times$4 triangulated Miura-like
		tessellation whilst (d), (e) and (f) are the converged results: (a)
		the base mesh formed on $\textbf{X}_{l}$ by the parametric coordinate
		lines $r=r_{i}$ and $s=s_{i}$ for $i=1,2,\cdots,9$; the $r_{i}$-coordinate
		lines are equispaced in the parametric coordinate $(\Delta r=1/4)$
		and so are the $s$-coordinate lines; 
		(b) the vertexes on the $s_{2i}$-coordinate
		lines ($i=1,2,3,4$) are moved along the $r$-coordinate by $\Delta r$;
		(c) the vertexes on the $r_{2i}$-coordinate line ($i=1,2,3,4$) are
		moved up to the upper target surface; (d) the converged folded form
		and $\gamma$ ($\approx77.1^\circ$) is
		the dihedral angle at the crease; (e) the folded form with the
		upper and lower target surfaces shown; (f) the crease pattern where
		the solid dots and squares are the vertexes attached to the
		lower and upper target surfaces, respectively.}
\end{figure*}

It is of practical interest that the origami, when partially folded,
the vertexes on each side approximate different given surfaces. Without
losing generality, the two surfaces are denoted as lower surface $\textbf{X}_{l}$
and upper surface $\textbf{X}_{u}$. The construction of the initial
triangulated Miura-like tessellation follows similar procedure as
that for the single target surface. For clarity, the case with $\textbf{X}_{l}=\{r,s,0\}$
and $\textbf{X}_{u}=\{r,s,(1+rs)/2\}$ is considered as an example.
Fig.\ref{fig:2surf_zxy_demo}(a) shows the base mesh generated on
the lower surface $\textbf{X}_{l}$ by the $r$- and $s$-coordinate
lines. The $r_{i}$-coordinate lines are equispaced in the parametric
coordinate $(\Delta r=1/4)$ and so are the $s$-coordinate lines.
The vertexes on the $s_{2i}$-coordinate lines ($i=1,2,3$ and 4)
are then moved along the $r$-coordinate by $\Delta r$ to form the
mesh shown in Fig.\ref{fig:2surf_zxy_demo}(b). After that, the vertexes
on the $r_{2i}$-coordinate lines with $i=1,2,3$ and 4 are moved
to the upper target surface $\textbf{X}_{u}$ by changing the $z$-coordinate
of each vertex to $(1+rs)/2$. The vertexes on the blue lines in Fig.\ref{fig:2surf_zxy_demo}(c)
are then on the upper target surface. The initial construction does
not constitute a generalized Miura-ori tessellation as the constraints,
i.e., quadrilateral facet planarity, developability and flat-foldability,
are not fulfilled in general.

For the cell shown in Fig.\ref{fig:GM_cell}(a), the vertexes \{1,
3, 7, 9\} are attached to the lower surface whilst the vertex-5 is
attached to the upper surface. For the attached vertexes, the associated
parametric coordinates are the variables whilst the other vertexes
own the Cartesian coordinates as the variables. The construction then
resorts to the optimization algorithm (the \emph{fmincon} routine
in the current implementation) to enforce the constraints on quadrilateral
facet planarity, developability and flat-foldability, i.e., Eq.(\ref{eq:quad_planar}),
(\ref{eq:tri_devep}) and (\ref{eq:tri_flat}), respectively. For
the case in Fig.\ref{fig:2surf_zxy_demo}, the \emph{fmincon} convergences
in 85 iterations and the final configuration is shown in Fig.\ref{fig:2surf_zxy_demo}(d)
and (e). The crease pattern is shown in Fig.\ref{fig:2surf_zxy_demo}(f).
It is checked that the constraints are fulfilled to machine accuracy.
Due to the Theorem 2 of the reference \cite{tachi2009generalization},
the pattern is rigid-foldable and no penetration is observed before
it is fully folded.

For a tessellation with $m\times n$ cells, the number of constraints
are the same as Eq.(\ref{eq:total_cons}). Compared to the single
target surface case, the central vertex (vertex-5 in Fig.) of each
cell is also attached to the upper surface, thus the total number
of variables and the number of excess DOFs are
\begin{equation}
N_{f}^{d}=N_{f}^{s}-mn=10mn+5m+5n+2\quad\text{and}\quad N_{f}^{d}-N_{c}=-2mn+9(m+n)\label{eq:excess_dof_d}
\end{equation}
respectively. For the case with $m=n$, it is required that $m\leq9$
to ensure that $N_{f}^{s}\geq N_{c}$, i.e., the number of variables
should be larger than that of the constraints. When $N_{f}^{s}<N_{c}$,
the optimization algorithm generally fails to find a generalized Miura-ori
tessellation.

\subsection{Further design examples\label{subsec:df_examples}}

\begin{figure*}
	\begin{centering}
		\includegraphics[width=14cm]{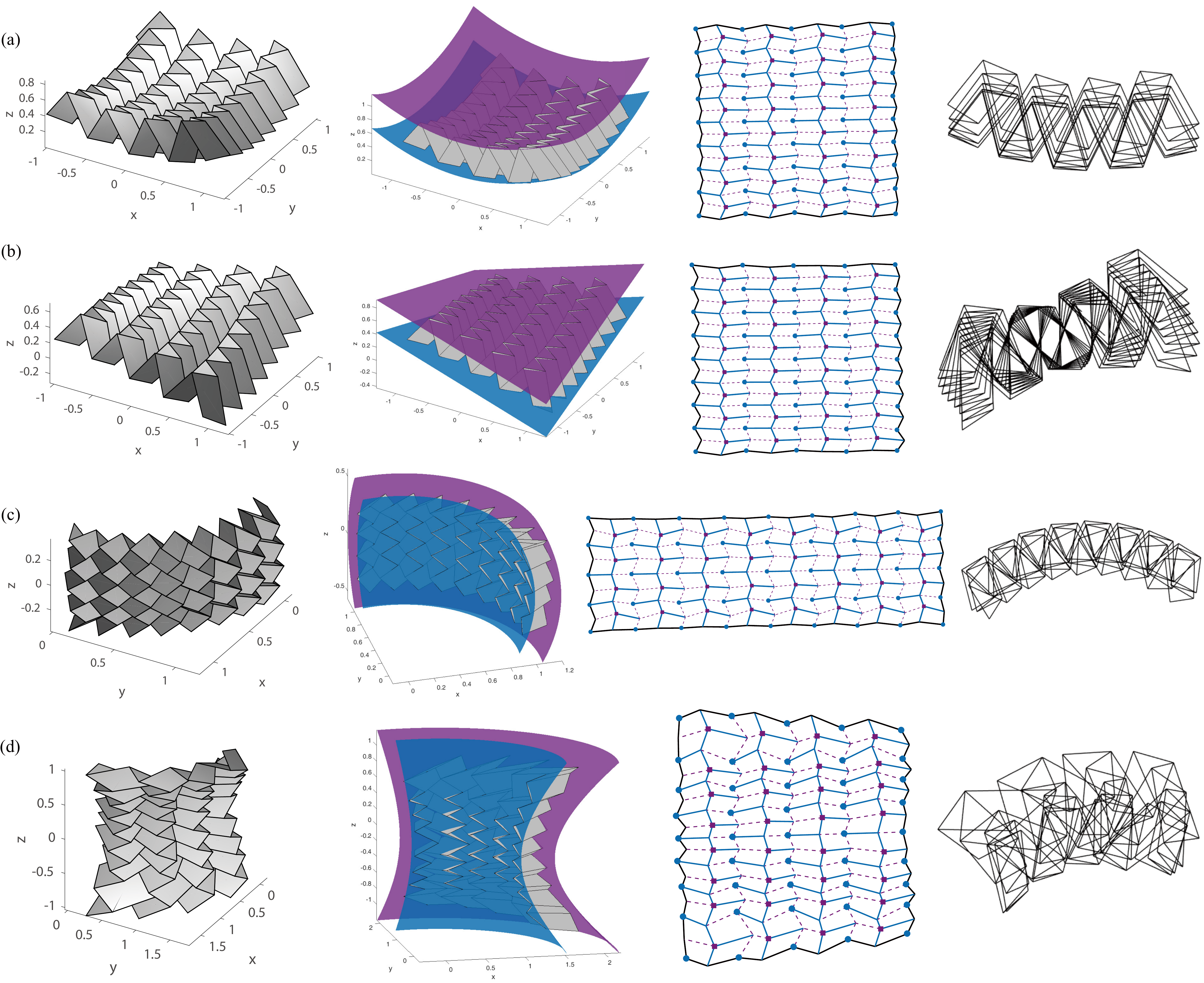}
		\par\end{centering}
	\caption{\label{fig:DGM_eaxples}
		Examples of approximating two target surfaces:
		the folded forms (first and second columns), the crease patterns (third
		column) and the fully folded state (fourth column), details are listed
		in Table \ref{tab:double_surf}. For figures in the second column,
		the lower and upper target surfaces are also shown.
		The solid dots and squares in the crease patterns are attached
		to the lower and upper surfaces, respectively.}
\end{figure*}

This subsection considers several examples to further illustrate the
design algorithm for approximating two target surfaces. The details
of the implementation are the same as those in section \ref{subsec:signle_de}.
Table \ref{tab:double_surf} summarizes the parameters for each cases.
The resulting folded forms are shown in the first and second columns
of Fig.\ref{fig:DGM_eaxples} whilst the crease patterns and the fully
folded states are listed in the third and fourth columns, respectively.

\begin{table*}
	\caption{\label{tab:double_surf}Summary of the design examples for approximating
		two target surfaces.}
	
	\centering{}\centerline{%
		\begin{tabular}{|c|c|c|c|c|c|}
			\hline 
			{\footnotesize{}cases} & {\footnotesize{}Lower target surface $\textbf{X}_{l}$} & {\footnotesize{}Upper target surface $\textbf{X}_{u}$} & {\footnotesize{}Cells} & {\footnotesize{}Iterations} & {\footnotesize{}Time(sec)}\tabularnewline
			\hline 
			\hline 
			{\footnotesize{}(a)} & {\footnotesize{}$\{r,s,-(r^{2}+s^{2})/5\}$} & {\footnotesize{}$\{r,s,-(r^{2}+s^{2})/5+1/2\}$} & {\footnotesize{}4$\times$8} & {\footnotesize{}129} & {\footnotesize{}28.1}\tabularnewline
			\hline 
			{\footnotesize{}(b)} & {\footnotesize{}$\{r,s,rs/4\}$} & {\footnotesize{}$\{r,s,rs/4+1/2\}$} & {\footnotesize{}4$\times$8} & {\footnotesize{}162} & {\footnotesize{}37.6}\tabularnewline
			\hline 
			{\footnotesize{}(c)} & {\footnotesize{}$\{\cos(s)\cos(r),\cos(s)\sin(r),\sin(s)\}$} & {\footnotesize{}$1.2\times\{\cos(s)\cos(r),\cos(s)\sin(r),\sin(s)\}$} & {\footnotesize{}8$\times$4} & {\footnotesize{}110} & {\footnotesize{}40.8}\tabularnewline
			\hline 
			{\footnotesize{}(d)} & {\footnotesize{}$\{\sqrt{1+s^{2}}\cos(r),\sqrt{1+s^{2}}\sin(r),s\}$} & {\footnotesize{}$\{\sqrt{2+s^{2}}\cos(r),\sqrt{2+s^{2}}\sin(r),s\}$} & {\footnotesize{}4$\times$8} & {\footnotesize{}84} & {\footnotesize{}24.3}\tabularnewline
			\hline 
	\end{tabular}}
\end{table*}

\subsection{Airfoil example\label{subsec:Airfoil-example}}

In this section, the lower and upper target surfaces are adopted from
the NACA-2412 airfoil, see Fig.\ref{fig:airfoil_geom}. The parametric
equations are given in the Appendix \ref{sec:append_airfoil}. It is difficult
to find explicit expressions regarding $\{x,y,z$\} for the target
surfaces. Thus, if the Cartesian coordinates of all the vertexes were
chosen as the variables, it would be inconvenient to enforce the attachment
of vertexes to the target surfaces through additional constraints
regarding the variables $\{x,y,z$\}. Using the parametric coordinates
for the pertinent vertexes as the variables avoids this difficulty.
Besides, the vertexes at the bottom and tip ends remain respectively
in the $y=0$ and $y=2$ planes during the optimization which is enforced
by adding linear constraints on the pertinent $y$-coordinate ($y=0$
for the bottom end and $y=2$ for the top end) or $s$-coordinate
for vertexes with parametric variables ($s=0$ for the bottom end
and $s=1$ for the top end).

\begin{figure*}
	\begin{centering}
		\centerline{\includegraphics[width=13cm]{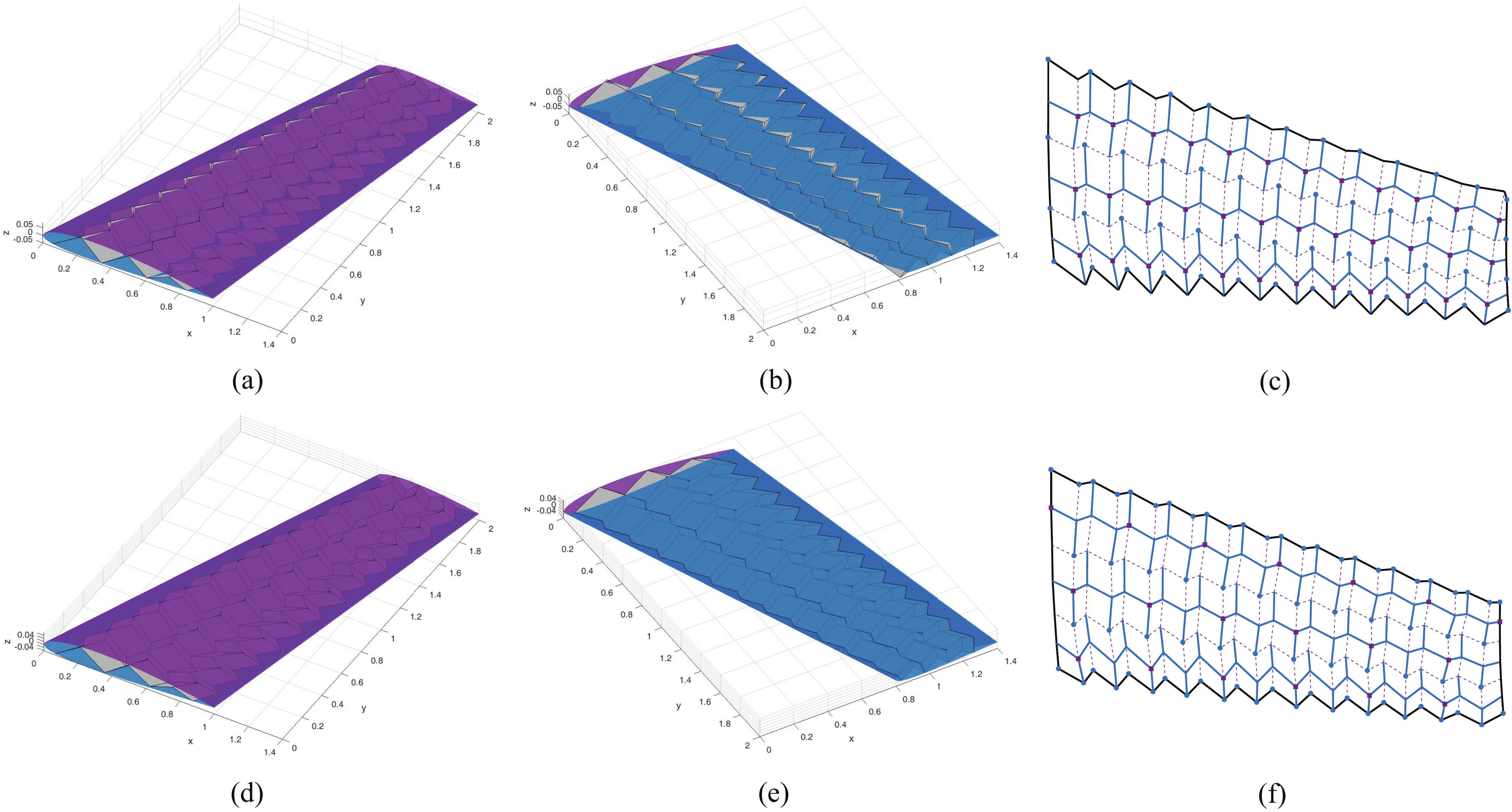}}
		\par\end{centering}
	\caption{\label{fig:airfoil_3x12}The resulting designs by attaching different
		sets of vertexes to the target surfaces: the attached vertexes are
		indicated by dots (on lower surface) and squares (on upper
		surface) in the crease patterns in (c) and (f); (a) and (b) are the
		folded form for (c) whilst (d) and (e) are the counterparts for (f).}
\end{figure*}

\begin{figure*}
	\begin{centering}
		\centerline{\includegraphics[width=13cm]{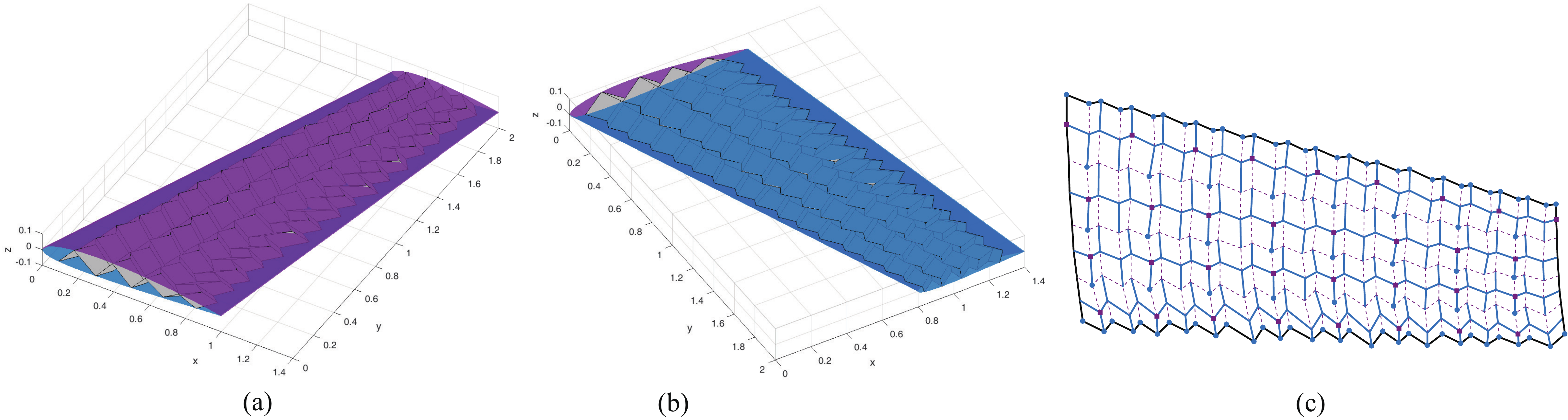}}
		\par\end{centering}
	\caption{\label{fig:airfoil_4x16}The designs with 4$\times$16 cells: (a)
		and (b) are the folded form; (c) shows the crease pattern where the
		solid dots and squares are attached to the lower and upper target
		surfaces, respectively.}
\end{figure*}

First, a tessellation with 3$\times$12 cells is considered. Similar
to the previous examples, the vertexes \{1,3,7,9\} and \{5\} of each
cell are, respectively, attached to the lower and upper target surfaces.
The resulting folded form is shown in Fig.\ref{fig:airfoil_3x12}(a)
and (b) with different view angles. Fig.\ref{fig:airfoil_3x12}(c)
shows the crease pattern in which the blue dots and purple squares
are on the lower and upper surfaces when the pattern is folded to
the target state. From Fig.\ref{fig:airfoil_3x12}(a) and (b), it
can be seen that a large number of vertexes (51 out of the 175) extrude
out of the target surface. By observing the relative position of the
vertexes on each zigzag line in Fig.\ref{fig:airfoil_3x12}(b) and
(c), the vertexes attached to the lower and upper target surfaces
are re-selected which is shown by the blue dots and purple squares
in Fig.\ref{fig:airfoil_3x12}(f). From the same initial triangulated
Miura-ori tessellation, the optimization algorithm yields the improved
folded form, see Fig.\ref{fig:airfoil_3x12}(d) and (e), and few vertexes
(10 out of the 175) are outside of the region. Next, the tessellation
with 4$\times$16 cells is considered. When the vertexes \{1,3,7,9\}
and \{5\} of each cell are, respectively, attached to the lower and
upper target surfaces, the algorithm fails to find a valid folded
form. This failure may indicate that too many vertexes are attached
to the target surfaces. To enlarge the solution space, fewer interior
vertexes are attached to the target surfaces, see Fig.\ref{fig:airfoil_4x16}(c).
The folded form is shown in Fig.\ref{fig:airfoil_4x16}(a) and (b).

\begin{figure}
	\begin{centering}
		\centerline{\includegraphics[width=8cm]{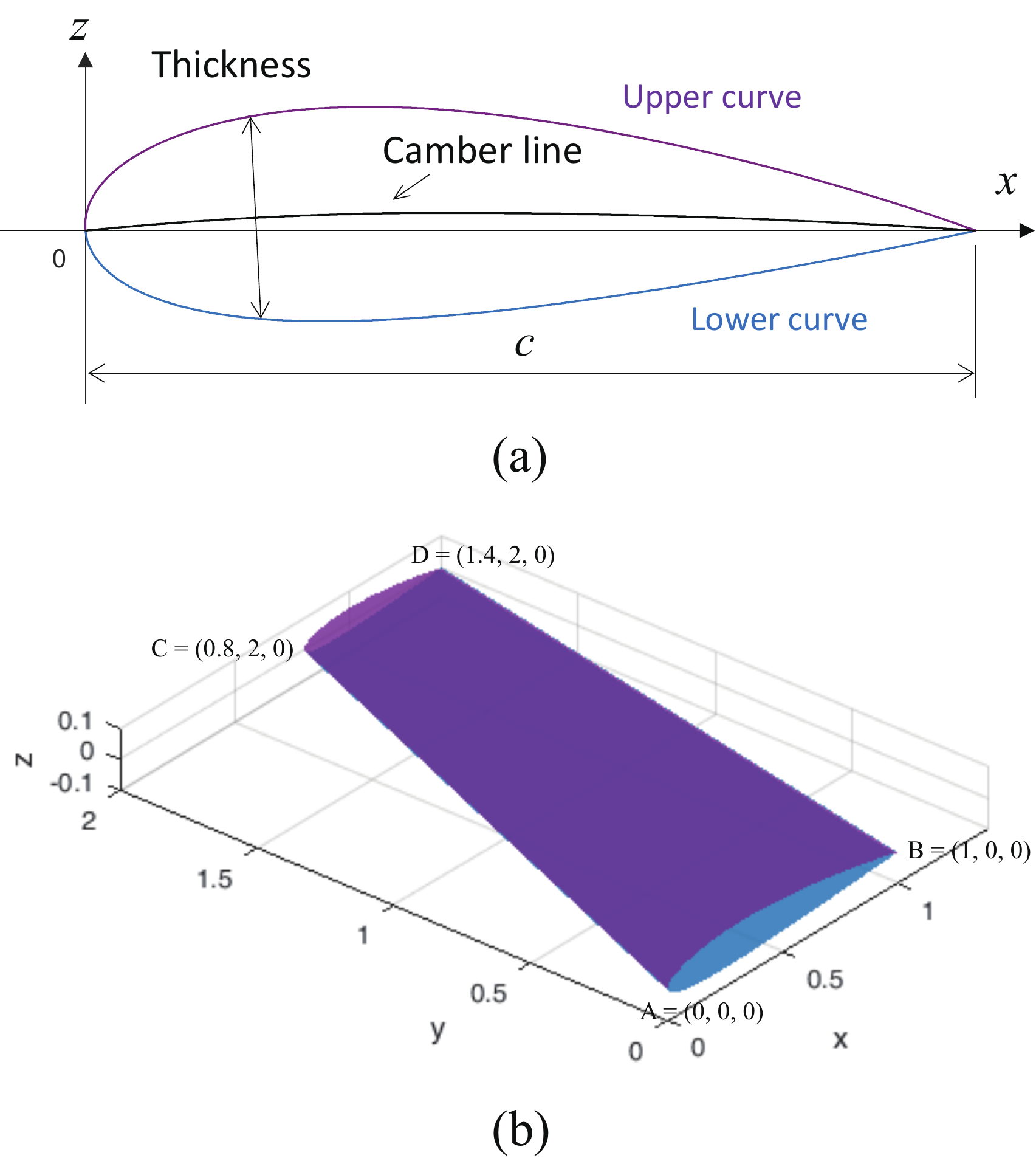}}
		\par\end{centering}
	\caption{\label{fig:airfoil_geom}
		(a) The cross section of the NACA four-digit
		airfoil \cite{moran2003introduction}; 
		(b) the lower and upper
		surfaces of the airfoil constructed from the cross sections
		at the bottom ($y=0$) and tip ($y=2$) end; A, B, C and D indicate
		the corner points.}
\end{figure}

\section{Conclusion}

This paper has presented the construction of rigid-foldable generalized
Miura-ori tessellations which can approximate curved parametric surfaces
based on constrained optimization. 
The algorithm is capable to approximate a single or double target surfaces by attaching portions of vertexes
to the pertinent target surfaces. All the inherent properties of standard
Miura-ori fold, including quadrilateral facet planarity, developability
and flat-foldability, are enforced through optimization. To enlarge
the searching space, the attached vertexes are allowed to slide on
the target surface through using their parametric coordinates as the
variables. Tessellations for target surfaces with varying curvatures
in both directions have been constructed which are rigid-foldable
with a single degree of freedom.

A limitation of the current algorithm is that the number of attached
vertexes is restricted such that the optimization problem is not over-constrained,
see Eq.(\ref{eq:excess_dof_s}) and (\ref{eq:excess_dof_d}) for the
approximation of single and double target surfaces, respectively.
While a re-selecting strategy is suggested and illustrated in Section
\ref{subsec:Airfoil-example}, vertexes to be attached to the target
surfaces are still chosen empirically in both layout and number. Neverthless,
the present algorithm can be an efficient tool to generate small,
rigid-foldable, curved patches which may be used as building blocks
to piece up larger rigid-folable crease patterns \cite{lang2017twists,lang2018rigidly,dieleman2019jigsaw}.
%It is noted that the level of approximation is not exactly controlled
%in the present algorithm which may be incorporated into the objective
%function in future works.

%%%%%%%%%%%%%%%%%%%%%%%%%%%%%%%%%%%%%%%%%%%%%%%%%%%%%%%%%%%%%%%%%%%%%%
\begin{acknowledgment}
The authors would like to thank Dr. Shuo Feng for insightful discussions.
This work is supported by the National Natural Science Foundation
of China (Nos. 11272303, 11072230, and 11802107), the Fundamental
Research Funds for the Central Universities of China (Nos. WK2480000001
and WK2090050040) and the support of Entrepreneurship and Innovation
Doctor Program in Jiangsu Province.
\end{acknowledgment}

\appendix

\section{Derivatives of the objective and constraint functions\label{sec:append_derivative}}

\begin{figure}
	\begin{centering}
		\includegraphics[width=6cm]{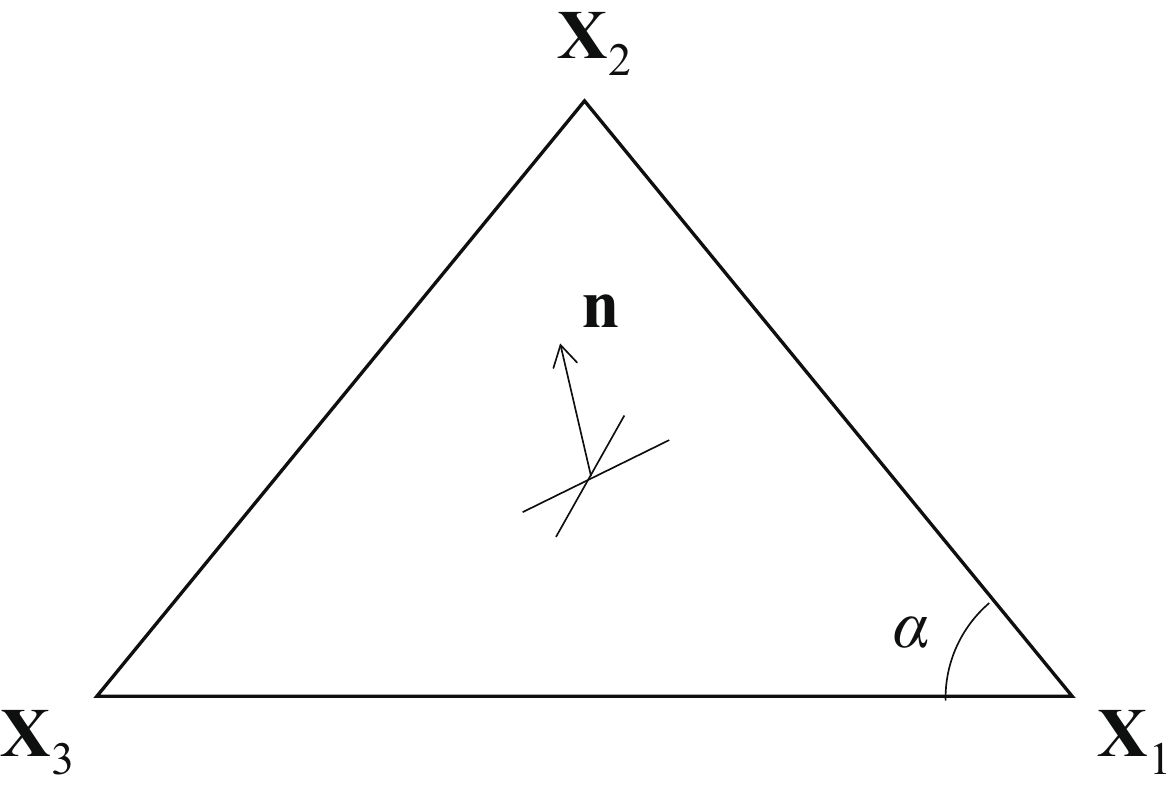}
		\par\end{centering}
	\caption{\label{fig:append_trig}A typical triangle and $\textbf{n}$ is its
		unit normal.}
\end{figure}

This appendix discusses the derivatives of the objective and constraint
functions which should be provided to enhance the efficiency of the
optimization solver. The derivatives with respective to (w.r.t) the
Cartesian coordinates are considered first. The derivatives of the
objective function in Eq.(\ref{eq:obj}) and the quadrilateral facet
planarity constraint in Eq.(\ref{eq:quad_planar}) w.r.t to the Cartesian
coordinates can be ready obtained and are not detailed here. For an
interior vertex, the developability and flat-foldability constraints
are the sum of the pertinent sector angles around the vertex. Each
sector angle is determined by the three vertexes of the triangle where
it resides. For a typical triangle shown in Fig.\ref{fig:append_trig},
the sector angle $\alpha$ is given by

\begin{equation}
\alpha=\arccos(\frac{(\textbf{X}_{2}-\textbf{X}_{1})\cdot(\textbf{X}_{3}-\textbf{X}_{1})}{\left\Vert \textbf{X}_{2}-\textbf{X}_{1}\right\Vert \left\Vert \textbf{X}_{3}-\textbf{X}_{1}\right\Vert })\ .
\end{equation}
The derivatives of $\alpha$ w.r.t the vertex coordinates are (also
see Eqs.(12), (13) and (14) of reference \cite{tachi2010freeform})

\begin{subequations}
	\begin{equation}
	\frac{\partial\alpha}{\partial\textbf{X}_{2}}=-\frac{\textbf{n}\times(\textbf{X}_{2}-\textbf{X}_{1})}{\left\Vert \textbf{X}_{2}-\textbf{X}_{1}\right\Vert ^{2}}\label{eq:dalp_dx2}
	\end{equation}
	
	\begin{equation}
	\frac{\partial\alpha}{\partial\textbf{X}_{3}}=\frac{\textbf{n}\times(\textbf{X}_{3}-\textbf{X}_{1})}{\left\Vert \textbf{X}_{3}-\textbf{X}_{1}\right\Vert ^{2}}\label{eq:dalp_dx3}
	\end{equation}
	
	\begin{equation}
	\frac{\partial\alpha}{\partial\textbf{X}_{1}}=-\frac{\partial\alpha}{\partial\textbf{X}_{2}}-\frac{\partial\alpha}{\partial\textbf{X}_{3}}\label{eq:dalp_dx1}
	\end{equation}
\end{subequations}
where $\textbf{n}=\frac{(\textbf{X}_{2}-\textbf{X}_{1})\times(\textbf{X}_{3}-\textbf{X}_{1})}{\left\Vert (\textbf{X}_{2}-\textbf{X}_{1})\times(\textbf{X}_{3}-\textbf{X}_{1})\right\Vert }$
is the unit normal of the triangle. Referring to the unit cell in
Fig.\ref{fig:GM_cell}, the derivatives of the developability and
flat-foldability constraints can then be obtained by adding the contributions
from the sector angles around the central vertex. The derivatives
w.r.t the parametric coordinates are obtained by the chain rule. For
instance, assume that the vertex-1 of the triangle in Fig.\ref{fig:append_trig}
is attached to the target surface and its unknown variables are $\{r_{1},s_{1}\}$,
then the derivatives of the sector angle w.r.t the parametric coordinates
are

\begin{equation}
\frac{\partial\alpha}{\partial r_{1}}=\frac{\partial\alpha}{\partial\textbf{X}_{1}}\cdot\frac{\partial\textbf{X}_{1}}{\partial r_{1}}\quad\text{and}\quad\frac{\partial\alpha}{\partial s_{1}}=\frac{\partial\alpha}{\partial\textbf{X}_{1}}\cdot\frac{\partial\textbf{X}_{1}}{\partial s_{1}}\ .
\end{equation}

\section{Check of self-intersection\label{sec:self-intersection}}

\begin{figure}
	\begin{centering}
		\includegraphics[width=6cm]{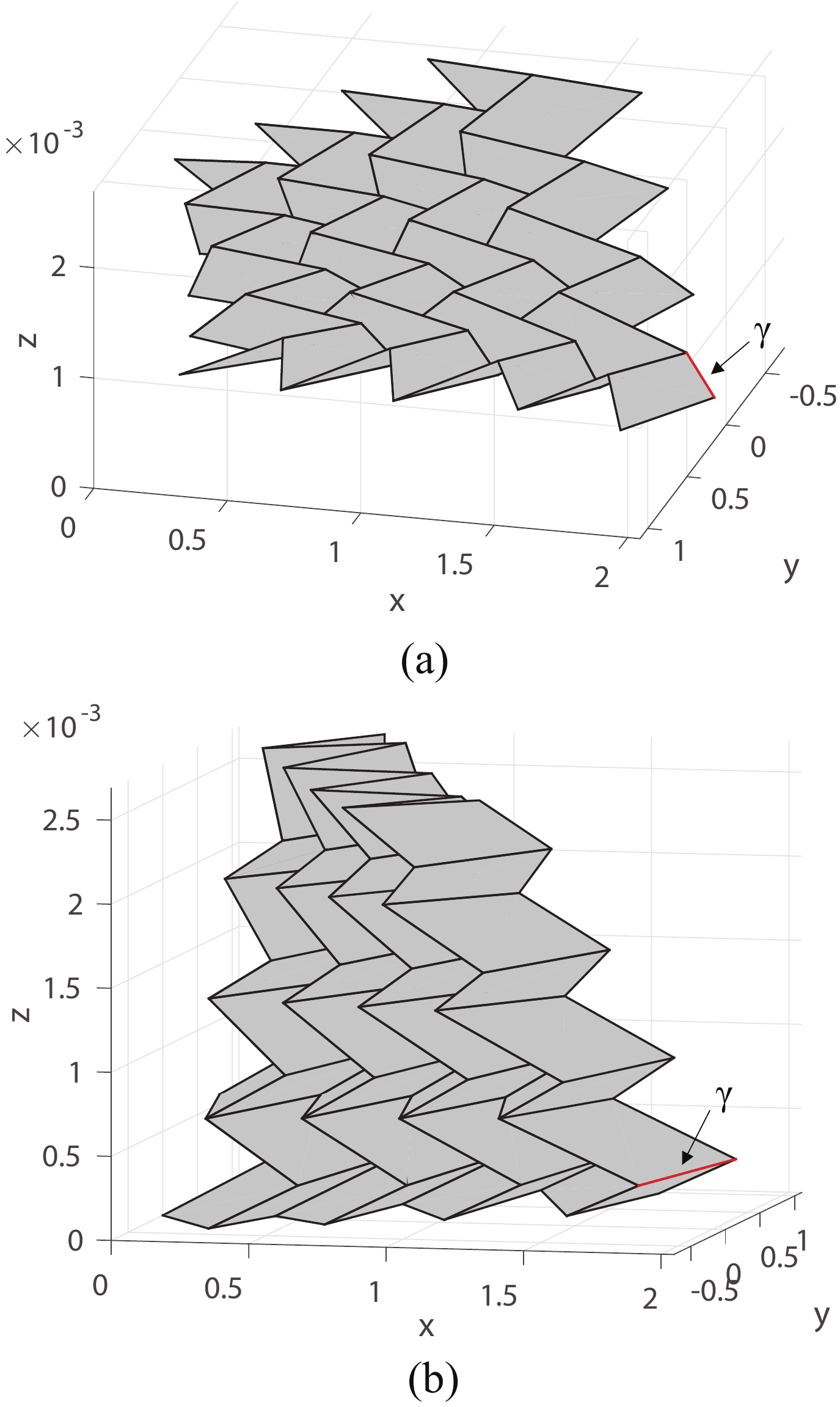}
		\par\end{centering}
	\caption{\label{fig:append_intersection}The folded form for the design resulf
		of $z=xy/2$ with $\gamma=0.1^{\circ}$: (a) and (b) are the folded
		form with different view angles. $\gamma$ is the dihedral angle at
		the crease indicated by the arrow. The scale of $z$-axis is much smaller than
		the others for better view.}
\end{figure}

Thanks to the Theorem 2 of the reference \cite{tachi2009generalization},
the crease pattern is rigid-foldable if an intermediate folded form
exists for the quadrilateral mesh consisting of developable and flat-foldable
vertexes. Nevertheless, the physically valid folded form of the crease
pattern is limited by self-intersection. For the generalized Miura-ori
pattern considered here, it is obvious that no self-intersection will
be observed when the crease pattern is folded only by a small amount.
As the folded form is of single degree-of-freedom (specified by the
characteristic fold angle $\gamma$) and the shape morphing is smooth,
contacts between the facets will happen when $\gamma$ reaches a certain
critical value. After the converged design has been obtained for each
case, we can view the three-dimensional folded form for given $\gamma$
and visually check whether global penetration occurs. For instance,
for the case of $z=xy/2$ in Section 2.1 and 2.2, the folded form
with $\gamma=0.1^{\circ}$ is plotted in Fig.\ref{fig:append_intersection}.
There is no penetration even for this almost fully folded state.

\section{Lower and upper surfaces of the airfoil\label{sec:append_airfoil}}

Fig.\ref{fig:airfoil_geom}(a) shows the cross section of the NACA
four-digit series airfoil in which $c$ is the airfoil chord. Introducing
the parametric coordinate $r=x/c$, the camber line and the thickness
distribution are, respectively, \cite{moran2003introduction}

\begin{subequations}
	\begin{equation}
	y_{c}(r)=\begin{cases}
	\begin{array}{c}
	\frac{\varepsilon cr}{p^{2}}(2p-r)\\
	\frac{\varepsilon c(1-r)}{(1-p)^{2}}(1+r-2p)
	\end{array} & \begin{array}{c}
	0\leq r<p\\
	p<r\leq1
	\end{array}\end{cases}\quad\text{and}
	\end{equation}
	
	\begin{equation}
	y_{t}(r)=10\tau c\left[0.2969\sqrt{r}-0.126r-0.3537r^{2}+0.2843r^{3}-0.1015r^{4}\right]
	\end{equation}
\end{subequations}
in which $\varepsilon$, $p$ and $\tau$ are parameters. For the
NACA-2412 airfoil considered in the example, the parameters are $\varepsilon=0.02,p=0.4$
and $\tau=0.12$. The lower and upper curves of the cross section
are respectively

\begin{equation}
\left\{ \begin{array}{c}
x_{L}(c,r)\\
z_{L}(c,r)
\end{array}\right\} =\left\{ \begin{array}{c}
c\times r+\frac{1}{2}y_{t}(r)\sin(\theta)\\
y_{c}-\frac{1}{2}y_{t}(r)\cos(\theta)
\end{array}\right\} 
\end{equation}
and
\begin{equation}
\left\{ \begin{array}{c}
x_{U}(c,r)\\
z_{U}(c,r)
\end{array}\right\} =\left\{ \begin{array}{c}
c\times r-\frac{1}{2}y_{t}(r)\sin(\theta)\\
y_{c}+\frac{1}{2}y_{t}(r)\cos(\theta)
\end{array}\right\} 
\end{equation}
where
\[
\theta=\arctan(\frac{dy_{c}}{dx})=\begin{cases}
\begin{array}{c}
\frac{2\varepsilon(p-r)}{p^{2}}\\
\frac{2\varepsilon(p-r)}{(1-p)^{2}}
\end{array} & \begin{array}{c}
0\leq r<p\\
p<r\leq1
\end{array}\end{cases}\ .
\]
The cross sections at the bottom ($y=0$) and tip ($y=2$) ends are
prescribed with the chord $c=1$ and $c=0.6$, respectively, see Fig.\ref{fig:airfoil_geom}(b).
The lower curves of the cross sections at the bottom and tip ends
are respectively
\begin{equation}
\textbf{X}_{L}^{b}=\left\{ \begin{array}{c}
x_{L}(1,r)\\
0\\
z_{L}(1,r)
\end{array}\right\} \quad\text{and}\quad\textbf{X}_{L}^{t}=\left\{ \begin{array}{c}
0.8+x_{L}(0.6,r)\\
2\\
z_{L}(0.6,r)
\end{array}\right\} \ .
\end{equation}
The lower surface of the airfoil is a ruled surface which is given
by

\begin{equation}
\textbf{X}_{L}(r,s)=(1-s)\textbf{X}_{L}^{b}+s\textbf{X}_{L}^{t}
\end{equation}
The upper surface of the airfoil is defined in a similar way, i.e.,

\begin{equation}
\textbf{X}_{U}(r,s)=(1-s)\textbf{X}_{U}^{b}+s\textbf{X}_{U}^{t}
\end{equation}
where
\begin{equation}
\textbf{X}_{U}^{b}=\left\{ \begin{array}{c}
x_{U}(1,r)\\
0\\
z_{U}(1,r)
\end{array}\right\} \quad\text{and}\quad\textbf{X}_{U}^{t}=\left\{ \begin{array}{c}
0.8+x_{U}(0.6,r)\\
2\\
z_{U}(0.6,r)
\end{array}\right\} \ .
\end{equation}
are the upper cross-sectional curves at the bottom and top ends, respectively.

%%%%%%%%%%%%%%%%%%%%%%%%%%%%%%%%%%%%%%%%%%%%%%%%%%%%%%%%%%%%%%%%%%%%%%
% The bibliography is stored in an external database file
% in the BibTeX format (file_name.bib).  The bibliography is
% created by the following command and it will appear in this
% position in the document. You may, of course, create your
% own bibliography by using thebibliography environment as in
%
% \begin{thebibliography}{12}
% ...
% \bibitem{itemreference} D. E. Knudsen.
% {\em 1966 World Bnus Almanac.}
% {Permafrost Press, Novosibirsk.}
% ...
% \end{thebibliography}

% Here's where you specify the bibliography style file.
% The full file name for the bibliography style file 
% used for an ASME paper is asmems4.bst.
\bibliographystyle{asmems4}

% Here's where you specify the bibliography database file.
% The full file name of the bibliography database for this
% article is asme2e.bib. The name for your database is up
% to you.
%\footnotesize{}
\bibliography{JMR-19-1365}

\begin{thebibliography}{10}

\bibitem{miura1970proposition}
Miura, K., 1970.
\newblock ``Proposition of pseudo-cylindrical concave polyhedral shells''.
\newblock In Proceedings of IASS Symposium on Folded Plates and Prismatic
  Structures.

\bibitem{miura1985method}
Miura, K., 1985.
\newblock ``Method of packaging and deployment of large membranes in space''.
\newblock {\em title The Institute of Space and Astronautical Science report,
  {\bf 618}}, p.~1.

\bibitem{miura1972zeta}
Miura, K., 1972.
\newblock ``Zeta-core sandwich-its concept and realization''.
\newblock {\em title ISAS report/Institute of Space and Aeronautical Science,
  University of Tokyo, {\bf 37}}(6), p.~137.

\bibitem{klett2011designing}
Klett, Y., and Drechsler, K., 2011.
\newblock ``Designing technical tessellations''.
\newblock In Origami 5: Fifth International Meeting of Origami Science,
  Mathematics, and Education, Taylor \& Francis Group, Singapore, pp.~305--322.

\bibitem{heimbs2013foldcore}
Heimbs, S., 2013.
\newblock ``Foldcore sandwich structures and their impact behaviour: an
  overview''.
\newblock In {\em Dynamic failure of composite and sandwich structures}.
  Springer, pp.~491--544.

\bibitem{ma2018origami}
Ma, J., Song, J., and Chen, Y., 2018.
\newblock ``An origami-inspired structure with graded stiffness''.
\newblock {\em International Journal of Mechanical Sciences, {\bf 136}},
  pp.~134--142.

\bibitem{wei2013geometric}
Wei, Z.~Y., Guo, Z.~V., Dudte, L., Liang, H.~Y., and Mahadevan, L., 2013.
\newblock ``Geometric mechanics of periodic pleated origami''.
\newblock {\em Physical review letters, {\bf 110}}(21), p.~215501.

\bibitem{schenk2013geometry}
Schenk, M., and Guest, S.~D., 2013.
\newblock ``Geometry of miura-folded metamaterials''.
\newblock {\em Proceedings of the National Academy of Sciences, {\bf 110}}(9),
  pp.~3276--3281.

\bibitem{silverberg2014using}
Silverberg, J.~L., Evans, A.~A., McLeod, L., Hayward, R.~C., Hull, T.,
  Santangelo, C.~D., and Cohen, I., 2014.
\newblock ``Using origami design principles to fold reprogrammable mechanical
  metamaterials''.
\newblock {\em science, {\bf 345}}(6197), pp.~647--650.

\bibitem{pratapa2018bloch}
Pratapa, P.~P., Suryanarayana, P., and Paulino, G.~H., 2018.
\newblock ``Bloch wave framework for structures with nonlocal interactions:
  Application to the design of origami acoustic metamaterials''.
\newblock {\em Journal of the Mechanics and Physics of Solids, {\bf 118}},
  pp.~115--132.

\bibitem{tachi2009generalization}
Tachi, T., 2009.
\newblock ``Generalization of rigid foldable quadrilateral mesh origami''.
\newblock In Symposium of the International Association for Shell and Spatial
  Structures (50th. 2009. Valencia). Evolution and Trends in Design, Analysis
  and Construction of Shell and Spatial Structures: Proceedings, Editorial
  Universitat Polit{\`e}cnica de Val{\`e}ncia.

\bibitem{dieleman2019jigsaw}
Dieleman, P., Vasmel, N., Waitukaitis, S., and van Hecke, M., 2019.
\newblock ``Jigsaw puzzle design of pluripotent origami''.
\newblock {\em Nature Physics}, pp.~1--6.

\bibitem{tachi2010freeform_quad}
Tachi, T., 2010.
\newblock ``Freeform rigid-foldable structure using bidirectionally
  flat-foldable planar quadrilateral mesh''.
\newblock {\em Advances in architectural geometry 2010}, pp.~87--102.

\bibitem{lang2017twists}
Lang, R.~J., 2017.
\newblock {\em Twists, Tilings, and Tessellations: Mathematical Methods for
  Geometric Origami}.
\newblock AK Peters/CRC Press.

\bibitem{tachi2010freeform}
Tachi, T., 2010.
\newblock ``Freeform variations of origami''.
\newblock {\em J. Geom. Graph, {\bf 14}}(2), pp.~203--215.

\bibitem{gattas2013miura}
Gattas, J.~M., Wu, W., and You, Z., 2013.
\newblock ``Miura-base rigid origami: parameterizations of first-level
  derivative and piecewise geometries''.
\newblock {\em Journal of Mechanical Design, {\bf 135}}(11), p.~111011.

\bibitem{lang2018rigidly}
Lang, R.~J., and Howell, L., 2018.
\newblock ``Rigidly foldable quadrilateral meshes from angle arrays''.
\newblock {\em Journal of Mechanisms and Robotics, {\bf 10}}(2), p.~021004.

\bibitem{zhou2015design}
Zhou, X., Wang, H., and You, Z., 2015.
\newblock ``Design of three-dimensional origami structures based on a vertex
  approach''.
\newblock {\em Proc. R. Soc. A, {\bf 471}}(2181), p.~20150407.

\bibitem{wang2016folding}
Wang, F., Gong, H., Chen, X., and Chen, C., 2016.
\newblock ``Folding to curved surfaces: A generalized design method and
  mechanics of origami-based cylindrical structures''.
\newblock {\em Scientific reports, {\bf 6}}, p.~33312.

\bibitem{song2017design}
Song, K., Zhou, X., Zang, S., Wang, H., and You, Z., 2017.
\newblock ``Design of rigid-foldable doubly curved origami tessellations based
  on trapezoidal crease patterns''.
\newblock {\em Proc. R. Soc. A, {\bf 473}}(2200), p.~20170016.

\bibitem{hu2019design}
Hu, Y.~C., Liang, H.~Y., and Duan, H.~L., 2019.
\newblock ``Design of cylindrical and axisymmetric origami structures based on
  generalized miura-ori cell''.
\newblock {\em Journal of Mechanisms and Robotics, {\bf 11}}(5), p.~051004.

\bibitem{dudte2016programming}
Dudte, L.~H., Vouga, E., Tachi, T., and Mahadevan, L., 2016.
\newblock ``Programming curvature using origami tessellations''.
\newblock {\em Nature materials, {\bf 15}}(5), p.~583.

\bibitem{waitukaitis2015origami}
Waitukaitis, S., Menaut, R., Chen, B. G.-g., and van Hecke, M., 2015.
\newblock ``Origami multistability: From single vertices to metasheets''.
\newblock {\em Physical review letters, {\bf 114}}(5), p.~055503.

\bibitem{bhooshan2016interactive}
Bhooshan, S., 2016.
\newblock ``Interactive design of curved-crease-folding''.
\newblock Master's thesis, University of Bath.

\bibitem{moran2003introduction}
Moran, J., 2003.
\newblock {\em An introduction to theoretical and computational aerodynamics}.
\newblock Courier Corporation.

\end{thebibliography}

% list of tables and figures
\newpage
\listoffigures
\listoftables

\newpage
\listofalgorithms
No algorithms.

\end{document}